\documentclass{ptephy_v1}%%%%%% to generate preprint number with ptep logo

\preprintnumber{XXXX-XXXX} %%% %%% Insert preprint number here

\usepackage{graphicx} %for dealing with figures.
\usepackage{bm}
\usepackage{Here}
\usepackage{mediabb}
\usepackage{booktabs}
\usepackage{multirow}

\begin{document}

\title{Investigation of kaonic atom optical potential by the high precision data of kaonic $^3$He and $^4$He atoms}

%%%% To generate auto affiliation numbers please use \author{}\affil{} command

\author{J. Yamagata-Sekihara}
\affil{Department of Physics, Kyoto Sangyo University, Kyoto 603-8555, Japan \email{yamagata@cc.kyoto-su.ac.jp}}

\author[2]{Y. Iizawa}
\affil[2]{Department of Physics, Tokyo Institute of Technology, 2-12-1 Ookayama, Meguro, Tokyo 152-8551, Japan}

\author[2]{D. Jido}

\author[3]{N. Ikeno}
\affil[3]{Department of Life and Environmental Agricultural Sciences, Tottori University, Tottori 680-8551, Japan}

\author[4,5]{T. Hashimoto}
\affil[4]{RIKEN Nishina Center for Accelerator-Based Science, RIKEN, Wako 351-0198, Japan}
\affil[5]{RIKEN Cluster for Pioneering Research, RIKEN, Wako 351-0198, Japan}

\author[6,5]{S. Okada} %%% Use optional bracket [3] to change the respective address
\affil[6]{Department of Mathematical and Physical Sciences, Chubu University, Kasugai, Aichi 487-8501, Japan}

\author[7]{S. Hirenzaki}
\affil[7]{Department of Physics, Nara Women's University, Nara 630-8506, Japan}

%%% To include the collaborator name... Please use the command "\collaborator"
%%% For example: \collaborator{ATLAS Collaboration}

\begin{abstract}%
We investigate kaonic atom optical potentials of $^3$He and $^4$He phenomenologically based on the latest high precision data of the $2p$ states of the kaonic atoms in helium (J-PARC E62).
We consider a simple phenomenological form for the optical potential proportional to the nuclear density distributions and clarify the meanings and implications of the data by determining the potential parameters by the latest data.
We find two sets of potential parameters for each of $K^--^3$He and $K^--^4$He that reproduce the data.
We then extract the potential parameters consistent with the data of $^3$He and $^4$He atoms simultaneously.
We report the possible strong isospin dependence of the potential, and the different strength of the data of $K^--^3$He and $K^--^4$He in the restrictions on the potential parameters.
We mention that the study of the $1s$ states of the kaonic helium atom is very interesting as a next step.
Observing these states could provide valuable information on potential parameters and kaonic nuclear $s$ states because of the mutual influence between the atomic and the nuclear states of kaon with the same orbital angular momentum.

%We mention that the study of the $1s$ states of the kaonic helium atom is very interesting as a next step, and the determination of the orbital angular momenta of the kaonic nuclear states could be helpful for the unified understandings of the atomic and nuclear bound states of kaon because of the mutual influence between the atomic and the nuclear states of kaon with the same orbital angular momentum.

\end{abstract}

\subjectindex{xxxx, xxx}

\maketitle

\section{Introduction\label{sec:1}}

Meson-nucleus systems are interesting research targets to investigate the in-medium meson properties and the aspect of the symmetry of the strong interaction at finite density~\cite{Batty:1997zp,Friedman:2007zza,Yamazaki:2012zza,Hayano:2008vn,Metag:2017yuh,Hirenzaki:2022dpt,Itahashi:2023boi,piAF:2022gvw}.
Especially, the ${\bar K}$-nucleus bound states, that is, kaonic atoms and kaonic nuclei, are promissing for the studies of the kaon properties at finite nuclear density and the kaon-nucleon interaction in the nuclear medium.
Since kaon is the lightest meson with strangeness and is believed to be one of the Nambu-Goldstone bosons of the SU(3) chiral symmetry breaking, the behavior of kaon at finite density is expected to provide quite relevant information on the studies of the strong interaction symmetry.
In addition, the quantitative knowledge of kaon properties at finite density is considered to be crucial to study the high density nuclear matter and to clarify the role of kaon there, for example, to show the clear criteria for the kaon condensation occurrence. 

Recently, there have been important developments in experimental studies of the kaon-nucleus bound states.
The experimental evidence of the kaonic nuclear states could open a new era of the meson-nuclear physics~\cite{J-PARCE15:2018zys}. 
And we are also interested in the recent experimental achievement of the high precision measurement of the kaonic $^3$He and $^4$He atoms~(J-PARC E62)~\cite{J-PARCE62:2022qnt}.
The accuracy of the data of the kaonic-helium atoms is almost two orders of magnitude better than the previous data and clearly opens up the precision frontier of the kaon physics. 
The experimental results clearly denied the existence of the possible large shifts and widths of the kaonic atoms suggested theoretically in connection with the studies of the kaonic nuclear states in Ref.~\cite{Akaishi:2005mw}.

Theoretically, the kaon-nucleus interaction has been studied using the large set of data of kaonic atoms by the $\chi^2$ fitting procedure~\cite{Batty:1997zp,Friedman:2007zza,Mares:2006vk}.
The developments of the chiral unitary approach \cite{Ramos:1999ku,Kaiser:1995eg,Oset:1997it} enable us to deepen our understandings of the hadron resonances in the ${\bar K}N$ channel \cite{Jido:2003cb} and kaon-nucleus bound systems \cite{Hirenzaki:2000da}. 
Global analyses of kaonic atom spectra~\cite{Cieply:2011yz,Cieply:2011fy,Friedman:2016rfd} suggested that repulsive shifts of kaonic atomic states stem from large nuclear absorption of kaon instead of the existence of nuclear bound states of kaon. 
Recently, the theoretical potentials have progressed to include higher order terms of density based on the models formulated on the basis of chiral symmetry~\cite{Cieply:2011yz,Cieply:2011fy,Friedman:2016rfd,Hrtankova:2019jky,Obertova:2022wpw}.
We also mention a recent work~\cite{Ichikawa:2020doq} in which kaon-nucleus interaction was extracted as a set of optical potential parameters by investigating observed $^{12}$C$(K^-,p)$ spectrum.
The progress has been also made in the studies of the kaonic nuclear states by the microscopic few body calculations~\cite{Akaishi:2002bg,Yamazaki:2002uh,Shevchenko:2007zz,Wycech:2008wf,Dote:2008hw,Ikeda:2008ub,Barnea:2012qa,Sekihara:2016vyd,Ohnishi:2017uni,Dote:2017wkk}.

In this article, we rather focus on kaonic atoms of specific nuclei, $^3$He and $^4$He, by investigating phenomenologically the kaonic-nucleus optical potentials based on the latest high precision data of the $2p$ states of the K$^--^{3,4}$He atoms~\cite{J-PARCE62:2022qnt}.
The high precision data by themselves enable us to constraint the optical potentials of individual nuclei and it is a good occasion to clarify the meanings and implications of the high precision data.
In Sect. 2, we explain our framework used in the present analyses and introduce the optical potential in a phenomenological form proportional to the nuclear density distributions.
In Sect. 3, we determine the potential parameters and summarize the features of the obtained potential.
We mention the restrictions on the potential parameters required by the data of $K^--^3$He and $K^--^4$He systems, and the possible strong isospin dependence of the optical potential.
We also show the comparison of the data with the calculated results by the theoretically proposed potentials~\cite{Friedman:2011np,Ramos:1999ku}.
We include a few additional discussions on the determined potential parameters in Sects. 3.4 and 3.5.
Section 4 is devoted to the summary and the conclusion of this article.
%In Appendix, we apply the potentials obtained in Sect. 3 to the various kaonic atoms and kaonic nuclei, and discuss the implications of these potentials to other kaonic systems.
In Appendix, we show the energy shifts and widths of various kaonic atoms with heavier nuclei as an application of the potentials obtained in Sect. 3 and see the possibility that the potentials provide nuclear states in the heavier nuclei with the same angular momentum as the atomic states.

\section{Theoretical model \label{sec:2}}

We adopt the standard optical potential description of the kaon-nucleus bound systems, and we solve the relativistic Klein-Gordon (KG) equation with the Lorentz scalar type optical potential $U(r)$ and the electromagnetic potential $V_{\rm em}(r)$, which is the time component of the Lorentz vector potential, to obtain the theoretical binding energies and widths of the kaonic atom states. The KG equation is written as, 
\begin{equation}
[-\nabla^2+\mu^2+2\mu U(r)]\phi(\bm{r})=[\omega-V_{\rm em}(r)]^2\phi(\bm{r})~,
\label{eq:KG}
\end{equation}
where $\mu$ is the kaon-nucleus reduced mass, $\omega$ the complex eigenenergy, and $\phi(\bm{r})$ the kaon wave function.
We write the complex energy with two real parameters as $\displaystyle \omega=E+\mu-i\frac{\Gamma}{2}$, where $E$ is the binding energy and $\Gamma$ is the width of the bound state.
As the electromagnetic potential~$V_{\rm em}(r)$, we consider the Coulomb potential between kaon and nucleus including the effects of the vacuum polarization.
The finite size charge distribution of the nucleus is also taken into account and $V_{\rm em}(r)$ is expressed as~\cite{Ikeno:2015ioa}, 
\begin{equation}
V_{\rm em}(r)=-\alpha\int\frac{\rho_{\rm ch}(r')Q(|\bm{r}-\bm{r'}|)}{|\bm{r}-\bm{r'}|}d\bm{r'}~,
\end{equation}
where $\alpha$ is the fine structure constant and $\rho_{\rm ch}(r)$ the charge distribution of the nucleus.
The function $Q$ is defined as,
\begin{equation}
Q(r)=1+\frac{2\alpha}{3\pi}\int^\infty_1~du~e^{-2mru}\left(1+\frac{1}{2u^2}\right)\frac{(u^2-1)^{1/2}}{u^2}~,
\end{equation}
where $m$ indicates the electron mass.
The second term on the right-hand side indicates the short range vacuum polarization effects.
We confirm that the electromagnetic potential adopted here is accurate enough to study the energy shifts due to the strong interaction reported in Ref.~\cite{J-PARCE62:2022qnt} by comparing the calculated results with those in Ref.~\cite{Santos:2004bw}.

We consider a  simple phenomenological optical potential $U(r)$, which is assumed to be proportional to the nuclear density distribution, defined as,
\begin{equation}
U(r) = (V_0+iW_0)\frac{\rho(r)}{\rho_0}~~,
\label{eq:1}
\end{equation}
where $\rho_0$ is the normal nuclear density $\rho_0=0.17~{\rm fm}^{-3}$.
The parameters $V_0$ and $W_0$ show the potential strength at the normal nuclear density, and are used as the parameters to study the kaon-nucleus bound systems in this article.
The nuclear density distribution $\rho(r)$ used in the optical potential is the distribution of the center of the nucleon.
The systematic analyses of the kaonic atoms using the similar potential have been reported in Ref.~\cite{Iizawa:2019uiu}.

\begin{figure}[htbp]
\centering
        \includegraphics[keepaspectratio, scale=0.35,angle=-90]{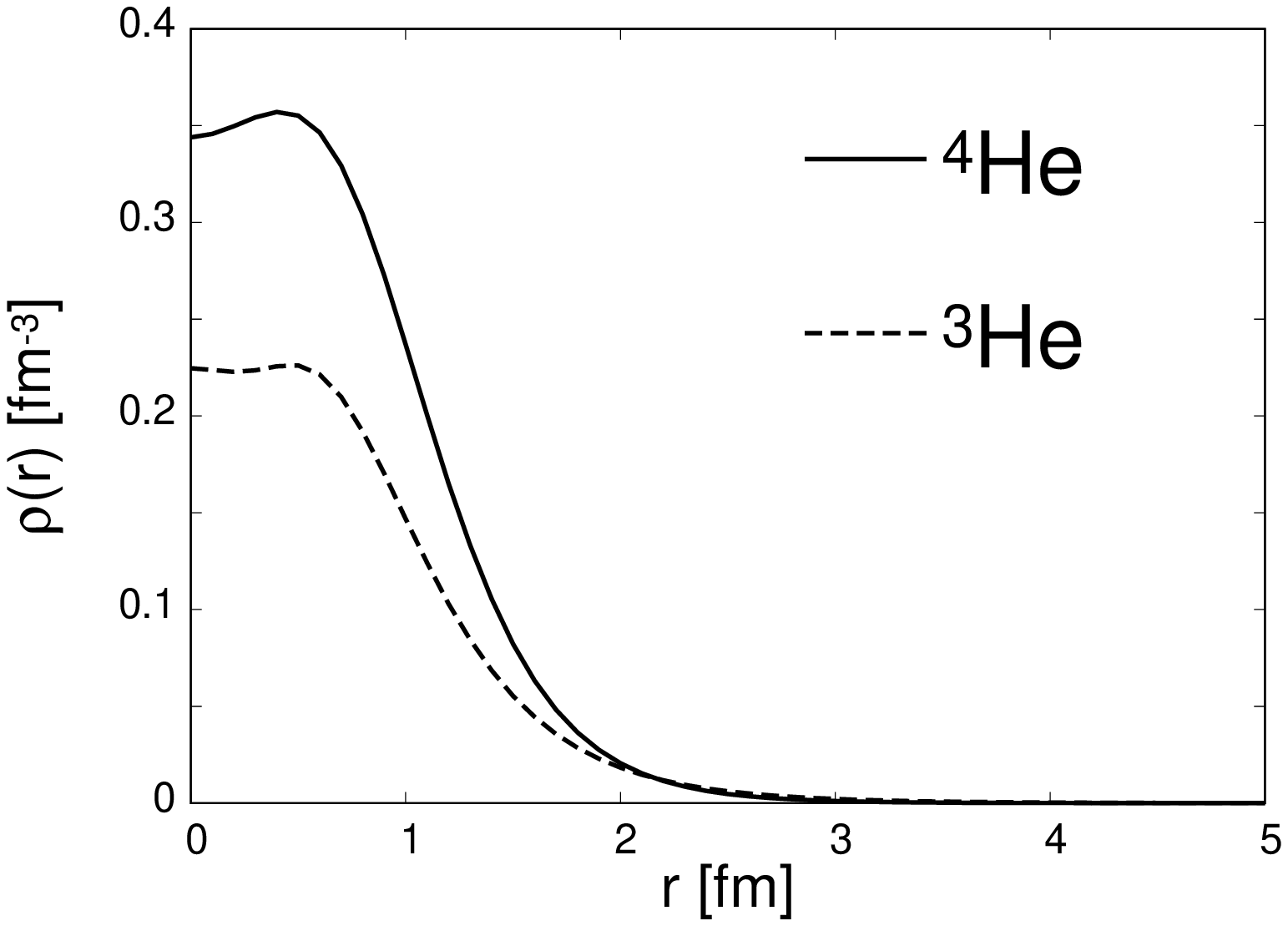}
\caption{The nuclear density distributions $\rho(r)$ of $^3$He and $^4$He in Eq.~(\ref{eq:1}) obtained by a microscopic calculation~\cite{Hiyama}.}
\label{Fig:11}
\end{figure}

We use realistic nuclear density distributions for $U(r)$ and realistic charge distributions for $V_{\rm em}(r)$, which are obtained by a precise few-body calculation~\cite{Hiyama} for $^3$He and $^4$He. The nuclear density distributions $\rho(r)$ are shown in Fig.~\ref{Fig:11}.
As can be seen in the figure, the central densities of these nuclei are higher than the normal nuclear density $\rho_0$, especially for $^4$He.
They are $\rho(0)\approx 1.32\rho_0$ for $^3$He and $\rho(0)\approx2.02\rho_0$ for $^4$He.
To use the realistic densities is important for the theoretical analyses here since the calculated results of the binding energies and the widths of the kaonic $2p$ atoms in He can be improved by a few eV from those obtained with the single Gaussian densities.

The potential picture for the kaonic atoms explained above is considered to be standard for the studies of the atomic states, however  this picture would be modified for the systems having so-called mesonic nuclear states where the degrees of freedom of each nucleon are expected to be more important and the few-body calculations of the whole systems would be required.

\section{Optical potential from $K^-- ^3$He and $K^-- ^4$He atomic $2p$ states\label{sec:3}}

\subsection{Experimental data and calculated results with theoretical potentials}
First, we briefly summarize the latest experimental data~\cite{J-PARCE62:2022qnt} and the calculated results by the existing theoretical potentials~\cite{Ramos:1999ku,Friedman:2011np} for $K^--^3$He and $K^--^4$He atoms.

The latest high precision experimental data reported in Ref.~\cite{J-PARCE62:2022qnt}, which we mainly study in this article, are summarized in Table~\ref{table:1}.
We mention here that the data have two notable features.
As far as we look at the central values of the data, we find (i) the opposite sign of the shifts of $K^--^3$He and $K^--^4$He, and (ii) the larger width of $K^--^3$He than $K^--^4$He.
These features could change in future because of the improvements of the errors of the present data.
The square root of the quadratic sum of the statistical and systematical errors is also shown in the parenthesis for each datum in Table~\ref{table:1}, and is considered to represent the experimental error of each datum in the following analyses of this article.
The transition energy shift $\Delta E$ is defined as $\Delta E=E^{\rm exp}_{3d\to 2p}-E^{\rm em}_{3d\to 2p}$ by the transition energies from $3d$ to $2p$ states observed in the experiment, $E^{\rm exp}_{3d\to 2p}$, and calculated with the electromagnetic potential, $E^{\rm em}_{3d\to 2p}$.
The positive shift $\Delta E>0$ means $E^{\rm exp}_{3d\to 2p}>E^{\rm em}_{3d\to 2p}$ and thus the larger experimental transition energy.
Since the observed energy $E^{\rm exp}_{3d\to 2p}$ is considered to correspond to the transition energy calculated with the full potential, $E^{\rm opt+em}_{3d\to 2p}$, the positive shift $\Delta E>0$ indicates that the effects of the optical potential makes the $2p$ state deeper than the corresponding pure electromagnetic state and that the attractive shift of the $2p$ state.
\begin{table}[H]
\caption{Experimental $3d\to 2p$ transition energy shifts and $2p$ widths of the kaonic $^3$He and $^4$He atoms~\cite{J-PARCE62:2022qnt} in units of eV.
The errors shown in the parentheses, which are the square root of the quadratic sum of the statistical and systematical errors in Ref.~\cite{J-PARCE62:2022qnt}, are considered to represent the experimental errors in this article.
It should be noted that the shift of $K^-- ^3$He is repulsive, while that of $K^-- ^4$He is attractive.}
\label{table:1}
\centering
\begin{tabular}{|c|cc|}
\hline
eV&Shift&Width\\
\hline
$K^--^3$He&$-0.2\pm0.4\,$(stat)$\pm0.3\,$(syst)&$2.5\pm1.0$\,(stat)$\pm0.4\,$(syst)\\
&($\pm0.5$)&($\pm1.1$)\\
\hline
$K^--^4$He&$0.2\pm0.3\,$(stat)$\pm0.2\,$(syst)&$1.0\pm0.6\,$(stat)$\pm0.3\,$(syst)\\
&$(\pm0.4)$&$(\pm0.7)$\\
\hline
\end{tabular}
\end{table}

\begin{figure}[htbp]
    \begin{tabular}{cc}
      \begin{minipage}[t]{0.5\hsize}
        \centering
        \includegraphics[keepaspectratio, scale=0.3,angle=-90]{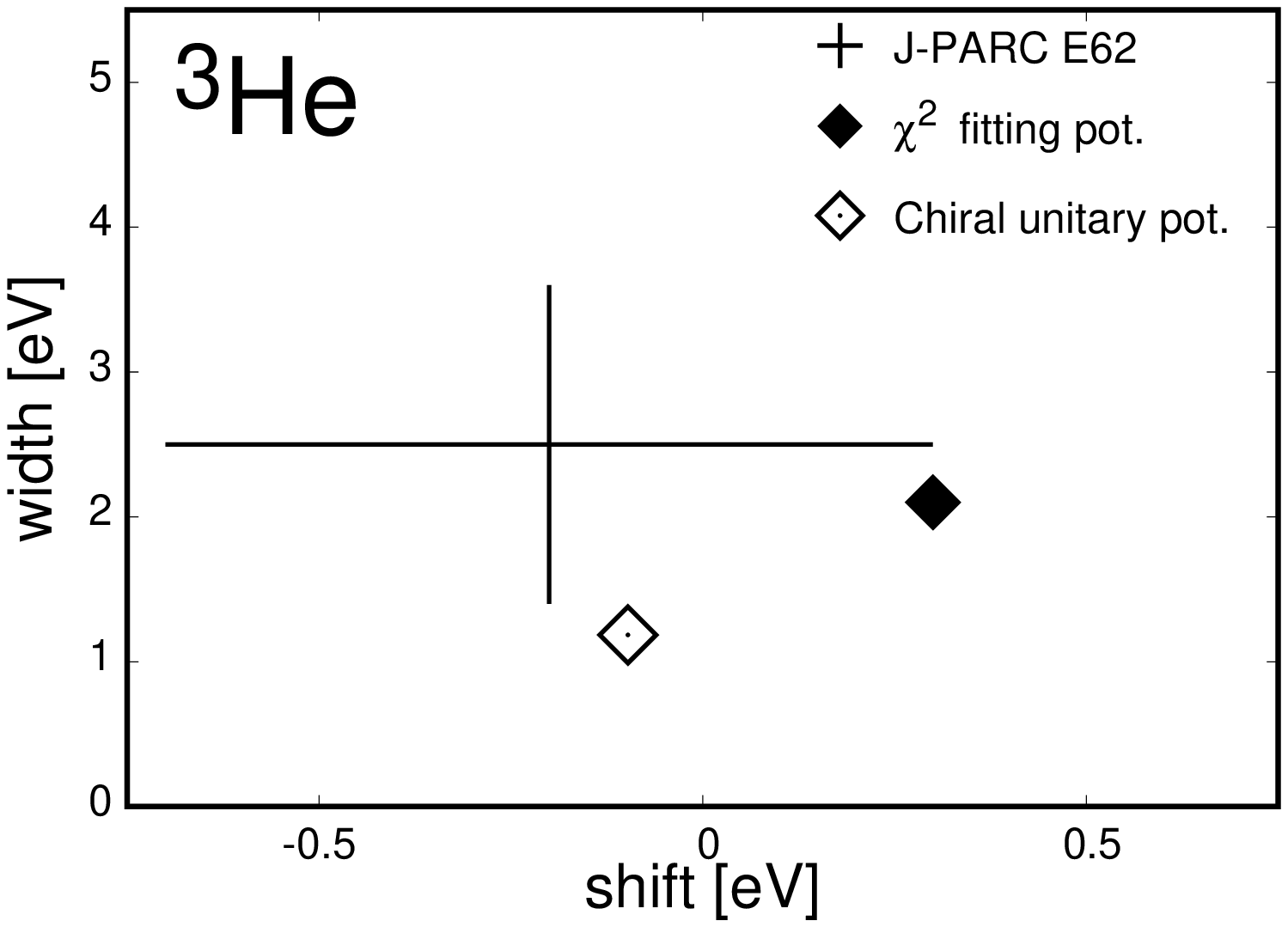}
%        \subcaption{Composite}
      \end{minipage} &
      \begin{minipage}[t]{0.5\hsize}
        \centering
        \includegraphics[keepaspectratio, scale=0.3,angle=-90]{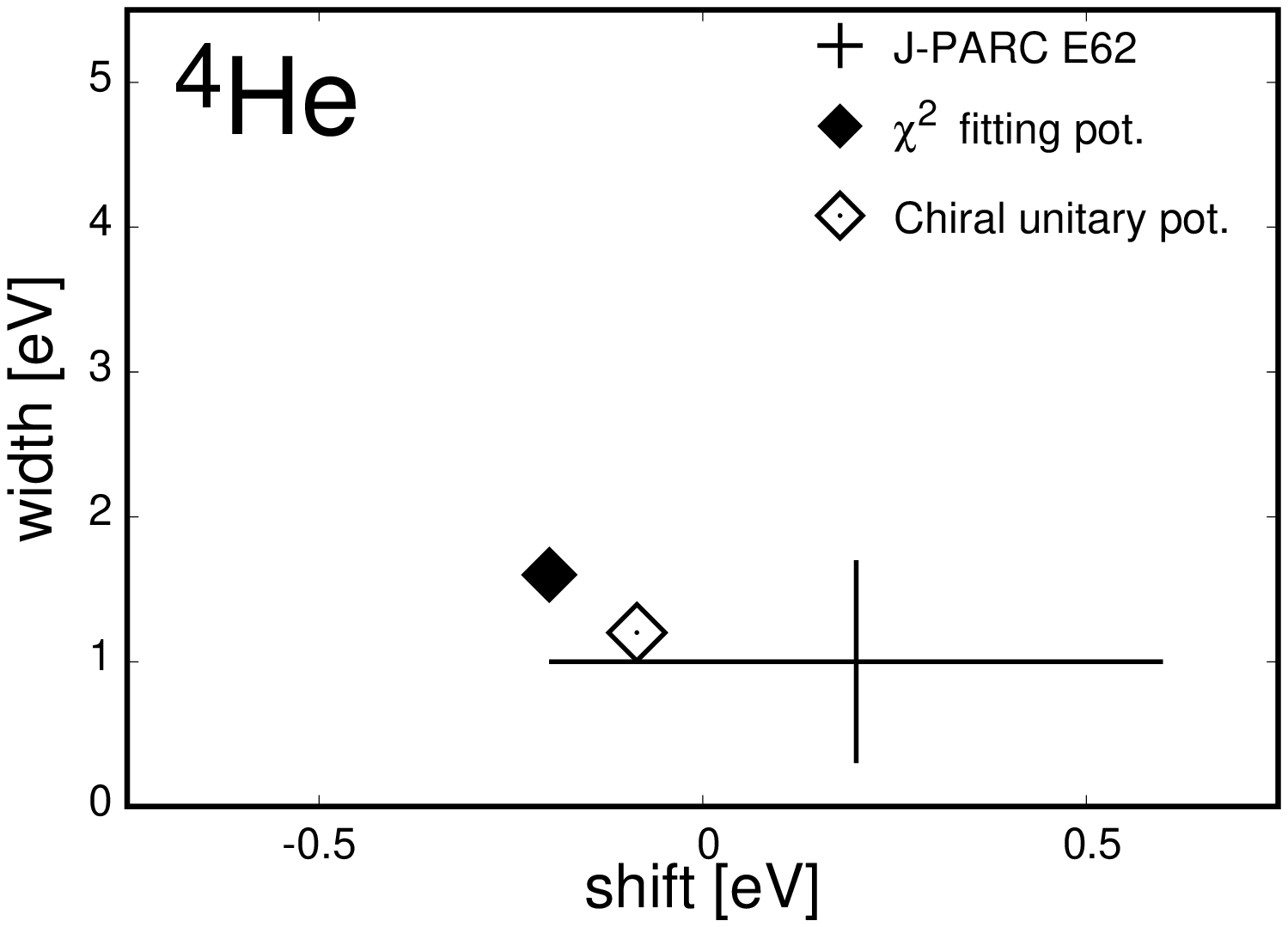}
  %      \subcaption{Gradation}
      \end{minipage} 
    \end{tabular}
     \caption{The calculated energy shifts of the $3d\to 2p$ transitions and the widths of the kaonic atom $2p$ states in $^3$He (Left) and $^4$He (Right) with the latest experimental data~\cite{J-PARCE62:2022qnt}.
     The theoretical results by the potential obtained by a large-scale fit ($\chi^2$ fitting pot.) are taken from Ref.~\cite{Friedman:2011np}.
     The theoretical potential in Ref.~\cite{Ramos:1999ku} is used for the chiral unitary model.}
             \label{Fig:4}
  \end{figure}

In order to check how well the existing theoretical potentials work for the new experimental data, we compare the latest data~\cite{J-PARCE62:2022qnt} with the energies of the atomic $2p$ and $3d$ states for $K^--^3$He and $K^--^4$He atoms calculated by theoretical potentials.
The potential obtained by a large-scale fit potential ($\chi^2$ fitting pot.) is taken from \cite{Friedman:2011np}.
Another potential in Ref.~\cite{Ramos:1999ku} obtained by the chiral unitary model is also used for the theoretical calculation.
This potential shows the strength ($-45,-61$) MeV at $\rho_0$.
The results are shown in Fig.~\ref{Fig:4} (Left) for $K^--^3$He and Fig.~\ref{Fig:4} (Right) for $K^--^4$He with the experimental data \cite{J-PARCE62:2022qnt}.
We find that both the potentials reproduce the data reasonably well.
However, the calculated width of $K^-- ^3$He by the chiral unitary potential is found to show small discrepancies with the experimental results.

It would be also interesting to see how the optical potential obtained in Ref.~\cite{Ichikawa:2020doq} works for the new data~\cite{J-PARCE62:2022qnt}.
In Ref.~\cite{Ichikawa:2020doq}, the parameters of the optical potential for $^{11}$B were extracted so as to reproduce the observed $^{12}$C($K^-,p$) spectrum for a wide range of the energy ($E-M_K$) from $-300$ MeV to $+40$ MeV and found to be ($-80, -40$) MeV at the nuclear center of $^{11}$B.
These correspond to ($V_0, W_0) = (-74, -37$) MeV for our formulation (4), which are the potential strength at the normal nuclear density.
This optical potential provides 0.46 eV for the shift and 1.01 eV for the width for the $2p$ state of the $K^--^4$He atom, which are consistent with the recent $^4$He data reported in Ref.~\cite{J-PARCE62:2022qnt} within the errors.

\subsection{Optical potential parameters of $^3$He and $^4$He fixed by the latest data}
\begin{figure}[htbp]
    \begin{tabular}{cc}   
      \begin{minipage}[t]{0.5\hsize}
        \centering
        \includegraphics[keepaspectratio, scale=0.28,angle=-90]{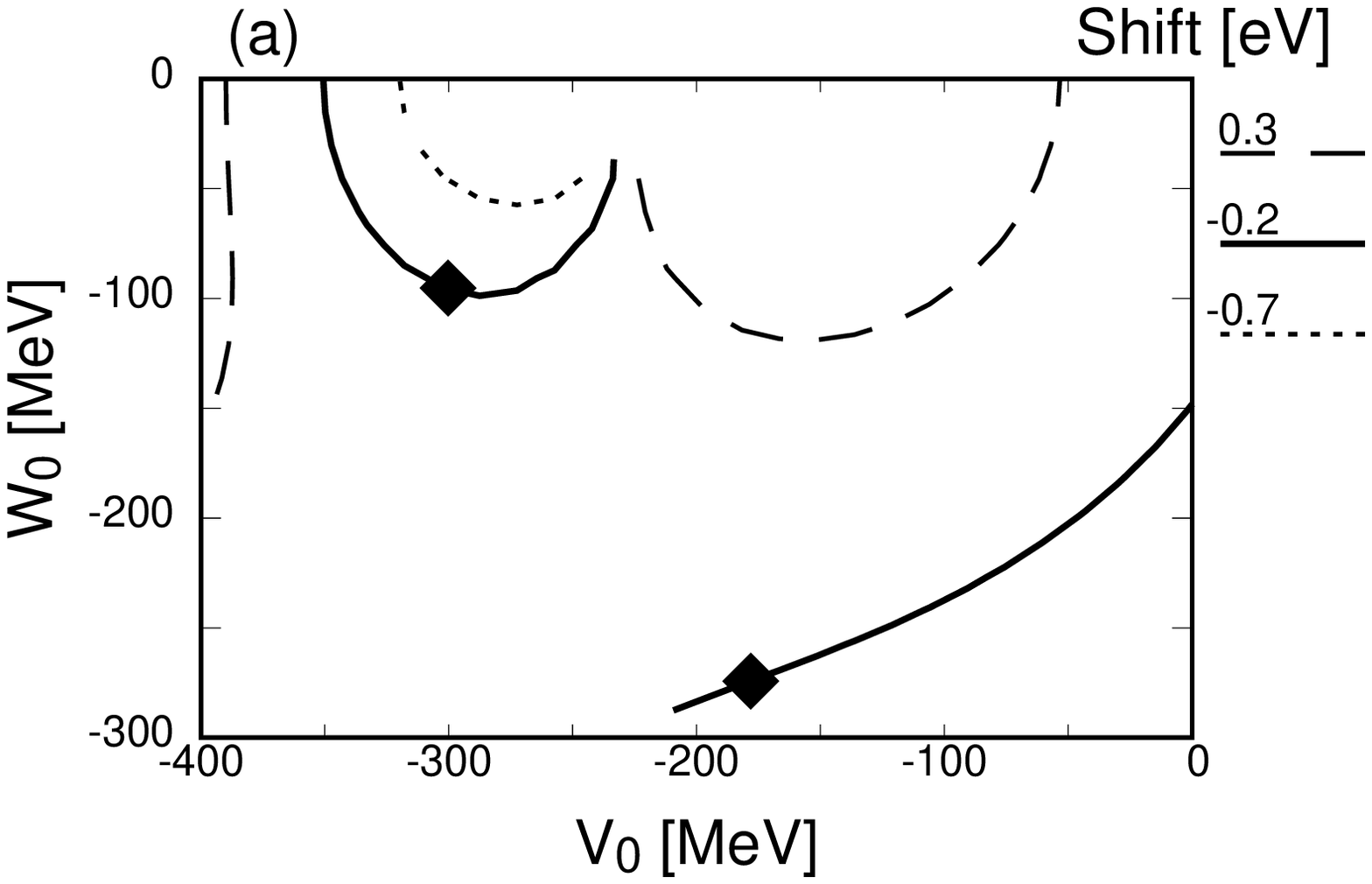}
     %   \subcaption{Fill}
      \end{minipage} &
      \begin{minipage}[t]{0.5\hsize}
        \centering
        \includegraphics[keepaspectratio, scale=0.28,angle=-90]{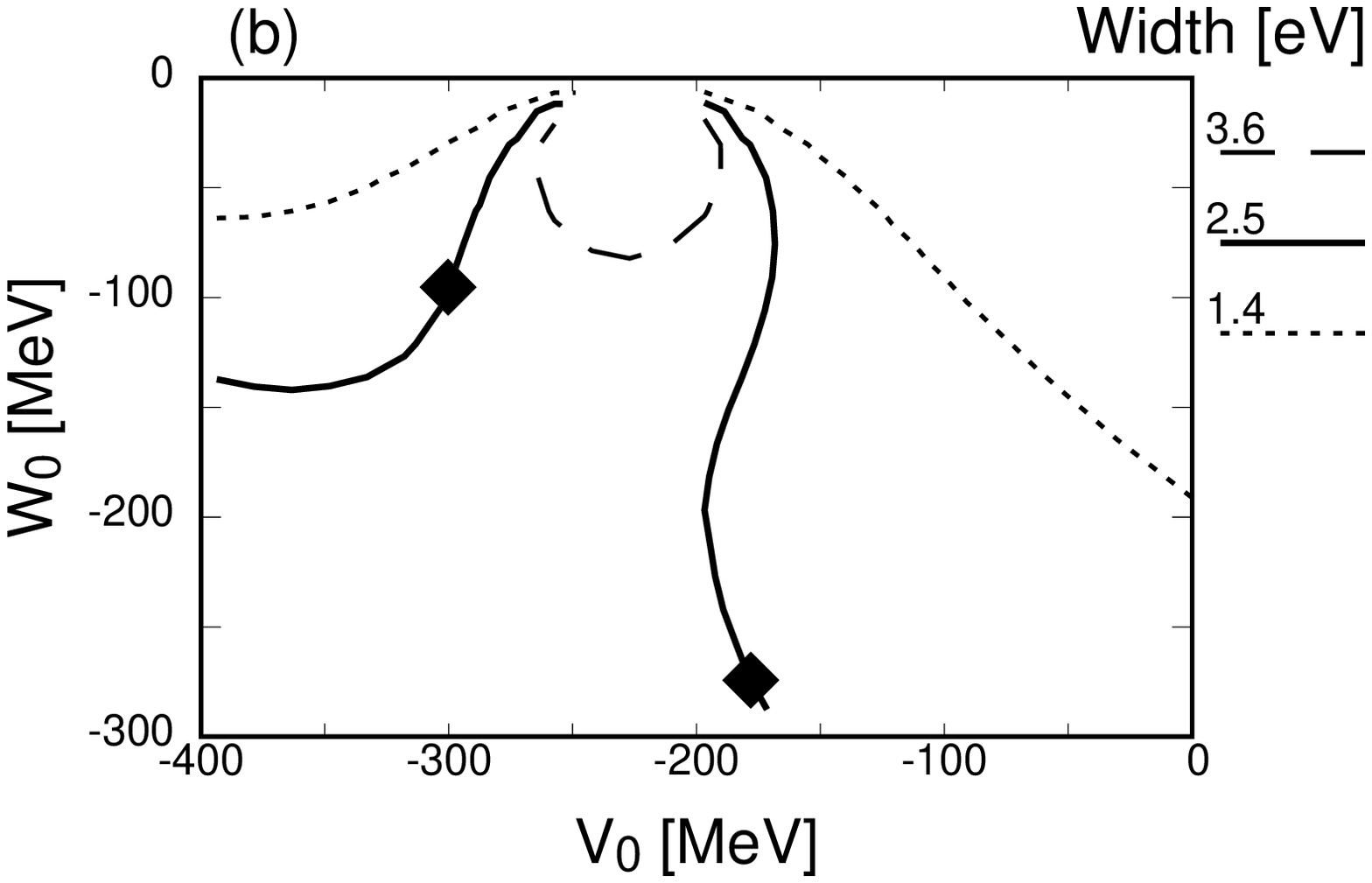}
        %\subcaption{Transform}
        \label{kore}
      \end{minipage}\vspace{-5mm} \\
      \multicolumn{2}{c}{
        \includegraphics[scale=0.3,angle=-90]{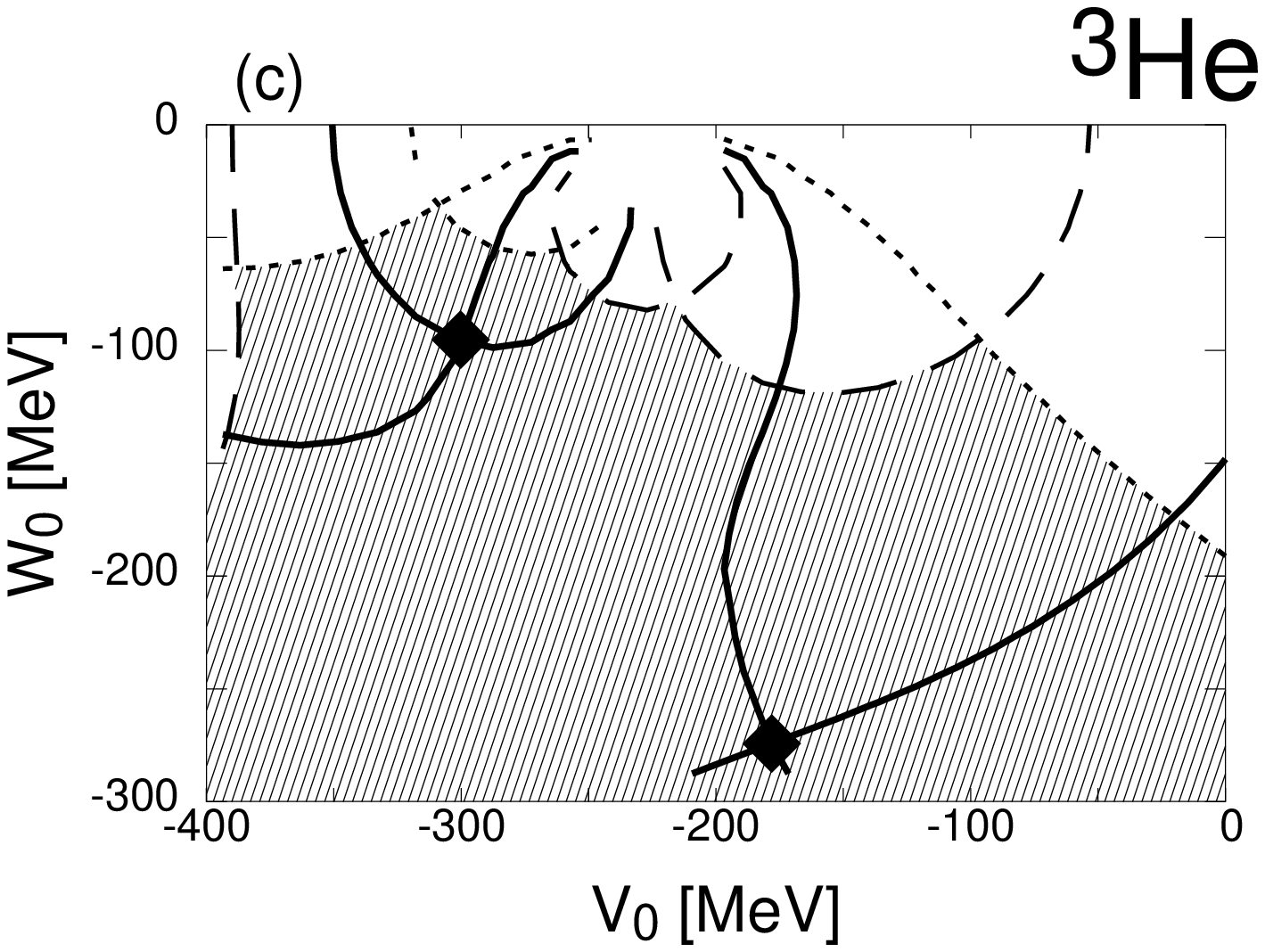}}
            \end{tabular}
     \caption{The contour plots of the shift of the $3d\to 2p$ transition energy (a)  and the $2p$ level width of the kaonic $^3$He atom (b) in the plane of the potential parameters $V_0$ and $W_0$ defined in Eq.~(\ref{eq:1}). The $V_0$ and $W_0$ values that reproduce the central values of the experimental data are shown by the solid lines. The $V_0$ and $W_0$ values reproducing the upper and lower bounds of the experimental data are shown by the dashed and the dotted lines. The experimental shift and width of the $K^--^3$He atom are reproduced simultaneously by the potential parameters $(V_0,W_0)=(-178,-274)$ MeV (HE3-A) and $(-300,-95)$ MeV (HE3-B) shown by the solid diamonds in the figures.
     Panel (c) shows the overlaid contour plot of the transition energy shift (a) and the width of the $K^--^3$He atom (b) in the same $V_0-W_0$ plane.
     The hatched region indicates the potential parameters that are consistent with the experimental data of the shift and width of $^3$He.}
     \label{Fig:1}
  \end{figure}

We extract here the kaon-nucleus optical potential parameters that are consistent with the latest data of the transition energy shifts and the level widths of the kaonic helium atoms in Ref.~\cite{J-PARCE62:2022qnt}.
We calculate the energy shift and width of kaonic $^3$He with given $V_0$ and $W_0$.
In Fig.~\ref{Fig:1}, we show the contour plots of the energy shifts and the widths of kaonic $^3$He in the $V_0-W_0$ plane of the potential parameters defined in Eq.~(\ref{eq:1}).
In Fig.~\ref{Fig:1} (a) and (b), the solid lines show the $V_0$ and $W_0$ values that reproduce the central values of the experimental data, and the dashed and dotted lines show the $V_0$ and $W_0$ values reproducing the upper and lower bounds of the experimental data, respectively.
We find that two sets of the potential parameters $(V_0,W_0)=(-178,-274)$ MeV~(HE3-A) and $(-300,-95)$ MeV (HE3-B) reproduce the both data of the shift and the width of the kaonic $^3$He atomic $2p$ state.
These solutions are listed in Table~\ref{table:1-5} and shown by the solid diamonds in Fig.~\ref{Fig:1}.
The obtained two parameter sets are summarized as a potential with a shallower real part and stronger absorption and one with a deeper real part and weaker absorption, respectively.
We present in Fig.~\ref{Fig:1} (c) the overlaid contour plot of the shift and the width shown in Figs.~\ref{Fig:1} (a) and (b).
The hatched area indicates the potential parameters that are consistent with the experimental data of the shift and width of $^3$He.
The relatively wide region of parameters are found to be acceptable for the kaonic $^3$He atomic data.

\begin{table}[!h]
\caption{The strengths of the optical potential defined in Eq.~(\ref{eq:1}) that reproduce the $2p$ atomic data in the $K^--^3$He system~\cite{J-PARCE62:2022qnt} listed in Table~\ref{table:1}.
The potential depths at normal nuclear density ($V_0$ and $W_0$ parameters) and those at the center of the $^3$He nucleus are shown for two sets of parameters HE3-A and HE3-B.
These potentials are shown as the solid diamonds in Figs.~\ref{Fig:1},~\ref{Fig:3},~\ref{Fig:11new},~\ref{Fig:10}.}
\label{table:1-5}
\centering
\begin{tabular}{|c|c|c||c|c|}
\hline
$^3$He&\multicolumn{2}{c||}{HE3-A}&\multicolumn{2}{c|}{HE3-B}\\\hline
[MeV]&Real&Imag&Real&Imag\\\hline
$U$ at $\rho=\rho_0$&\multirow{2}{*}{$-178$}&\multirow{2}{*}{$-274$}&\multirow{2}{*}{$-300$}&\multirow{2}{*}{$-95$}\\
($V_0$ and $W_0$)&&&&\\
\hline
\multirow{2}{*}{$U(r=0)$}&\multirow{2}{*}{$-235$}&\multirow{2}{*}{$-362$}&\multirow{2}{*}{$-396$}&\multirow{2}{*}{$-126$}\\
&&&&\\
\hline
\end{tabular}
\end{table}

\begin{figure}[htbp]
    \begin{tabular}{cc}   
      \begin{minipage}[t]{0.5\hsize}
        \centering
        \includegraphics[keepaspectratio, scale=0.28,angle=-90]{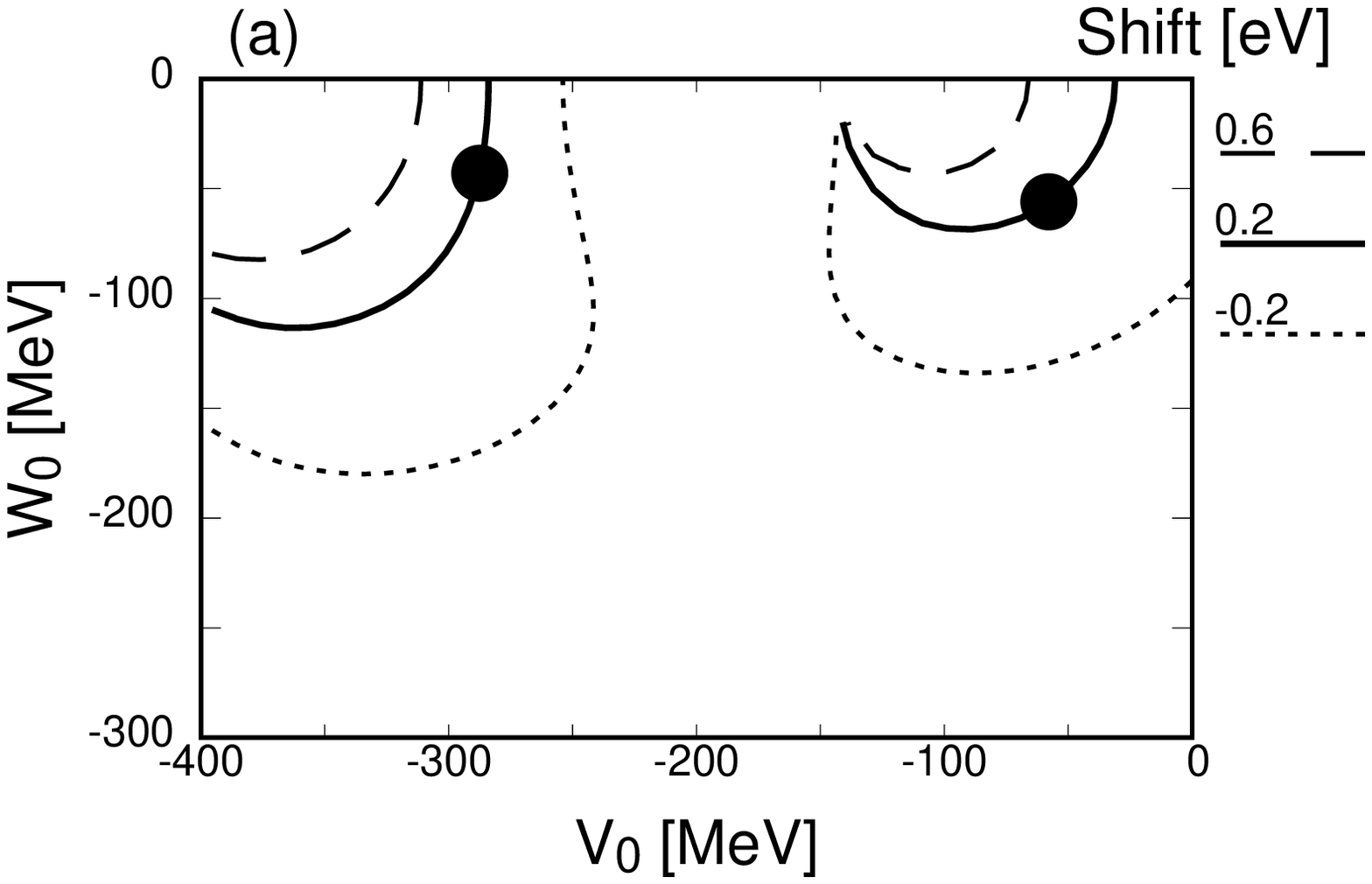}
     %   \subcaption{Fill}
      \end{minipage} &
      \begin{minipage}[t]{0.5\hsize}
        \centering
        \includegraphics[keepaspectratio, scale=0.28,angle=-90]{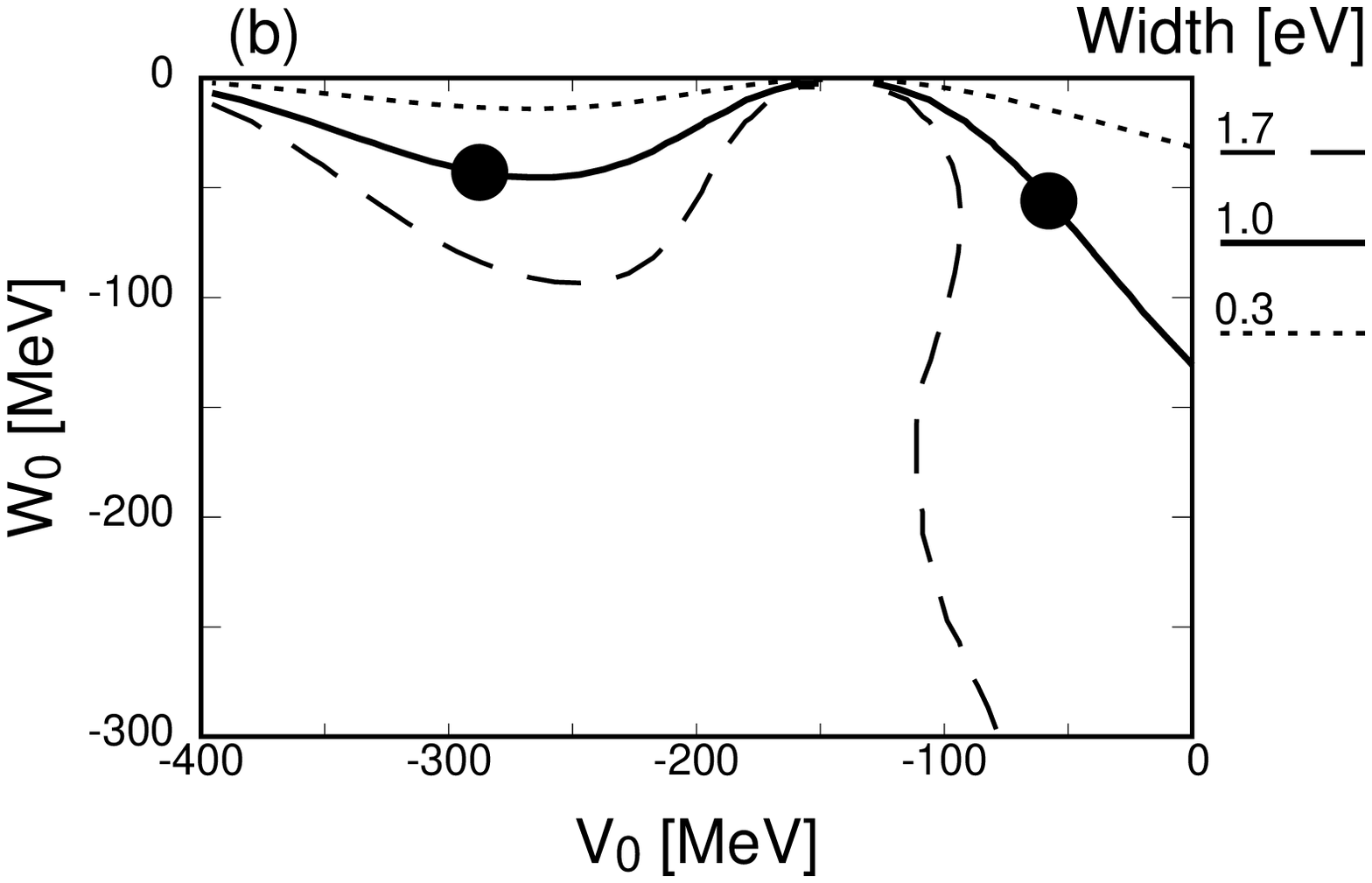}
        %\subcaption{Transform}
      \end{minipage} \\
      \multicolumn{2}{c}{
        \includegraphics[scale=0.3,angle=-90]{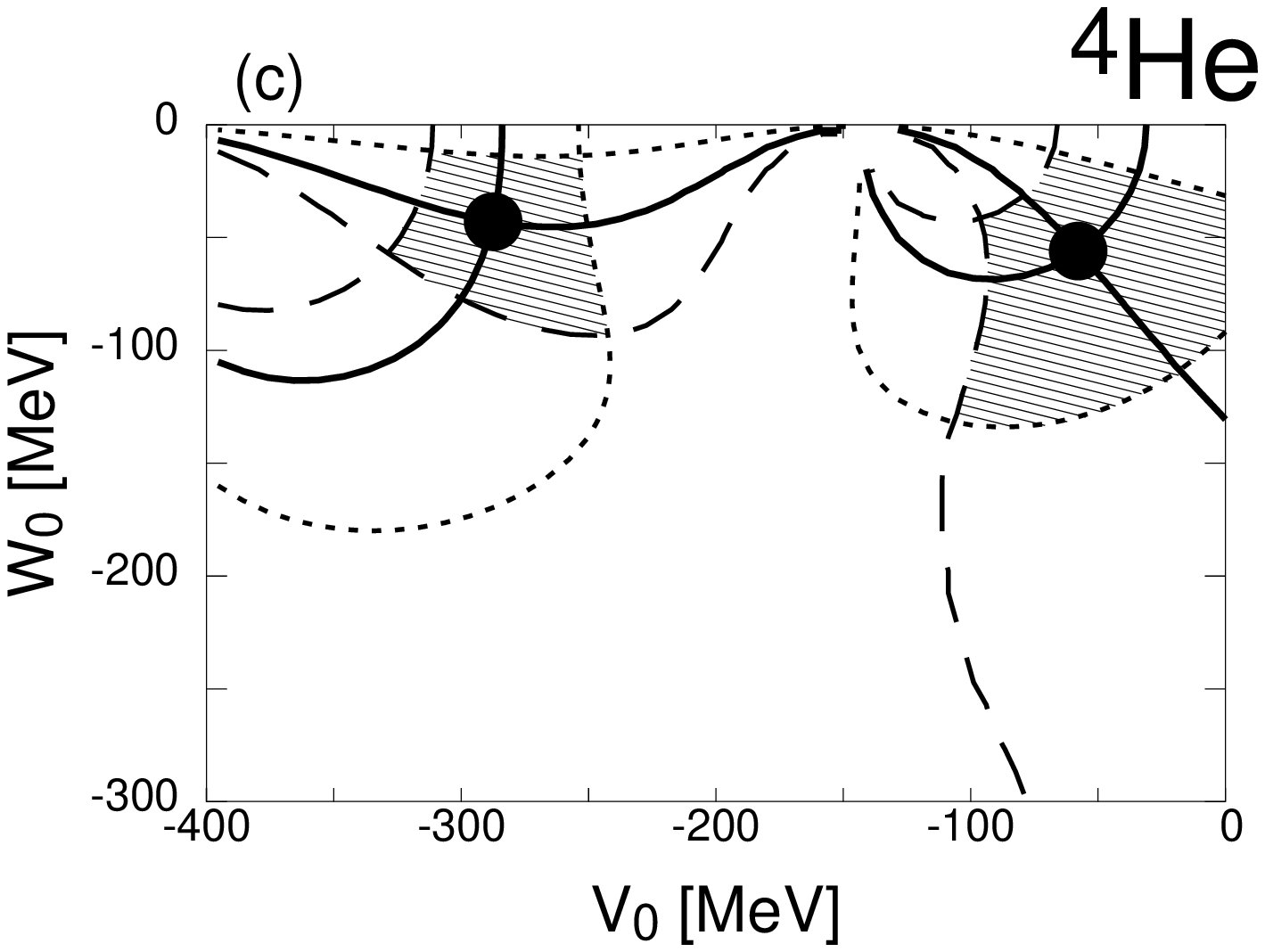}}
    \end{tabular}
     \caption{Same as Fig.~\ref{Fig:1} except for the kaonic $^4$He atomic state. The experimental shift and width of the $K^--^4$He atom are reproduced simultaneously by the potential parameters $(V_0,W_0)=(-58,-56)$ MeV (HE4-A) and $(-287,-43)$ MeV (HE4-B) shown by the solid circles in the figures.}
             \label{Fig:2}
  \end{figure}

In Fig.~\ref{Fig:2}, we show the results for the $K^--^4$He $2p$ atomic state obtained by the optical potential as in Fig.~\ref{Fig:1}.
For the $K^--^4$He system, we also find that two sets of the potential parameters $(V_0,W_0)=(-58,-56)$ MeV (HE4-A) and $(-287,-43)$ MeV (HE4-B) reproduce the central values of the experimental data both for shift and width of $2p$ state in $K^--^4$He.
These two sets of the potential parameters are listed in Table~\ref{table:1-7} and shown by the solid circles in Fig.~\ref{Fig:2}.
In this $K^--^4$He case, both parameter sets indicate the small values of the $W_0$ parameter for the imaginary potential, while the corresponding real potential parameters $V_0$ take small and large values.
It should be noted here that the central density $\rho(0)$ of $^4$He is close to $2\rho_0$ and 1.53 times larger than that of $^3$He.
In Fig.~\ref{Fig:2} (c), the overlaid contour plot of the shift and width of the $K^--^4$He atom are shown.
We find that the hatched areas, where the potential parameters are consistent with the data, exist in two regions in the $V_0-W_0$ plane separately.
The data of the $K^--^4$He atom provide the stronger constraints to the potential parameters than those of the $K^--^3$He atom.
\begin{table}[!h]
\caption{Same as Table~\ref{table:1-5} except for the $K^--^4$He system.
These potential parameters HE4-A and HE4-B are shown as the solid circles in Figs.~\ref{Fig:2},~\ref{Fig:3},~\ref{Fig:12},~\ref{Fig:10}.}
\label{table:1-7}
\centering
\begin{tabular}{|c|c|c||c|c|}
\hline
$^4$He&\multicolumn{2}{c||}{HE4-A}&\multicolumn{2}{c|}{HE4-B}\\\hline
[MeV]&Real&Imag&Real&Imag\\\hline
$U$ at $\rho=\rho_0$&\multirow{2}{*}{$-58$}&\multirow{2}{*}{$-56$}&\multirow{2}{*}{$-287$}&\multirow{2}{*}{$-43$}\\
($V_0$ and $W_0$)&&&&\\
\hline
\multirow{2}{*}{$U(r=0)$}&\multirow{2}{*}{$-117$}&\multirow{2}{*}{$-113$}&\multirow{2}{*}{$-581$}&\multirow{2}{*}{$-87$}\\
&&&&\\
\hline
\end{tabular}
\end{table}

\subsection{Combined study of kaonic $^3$He and $^4$He atoms}
\begin{figure}[htbp]
\centering
        \includegraphics[keepaspectratio, scale=0.35,angle=-90]{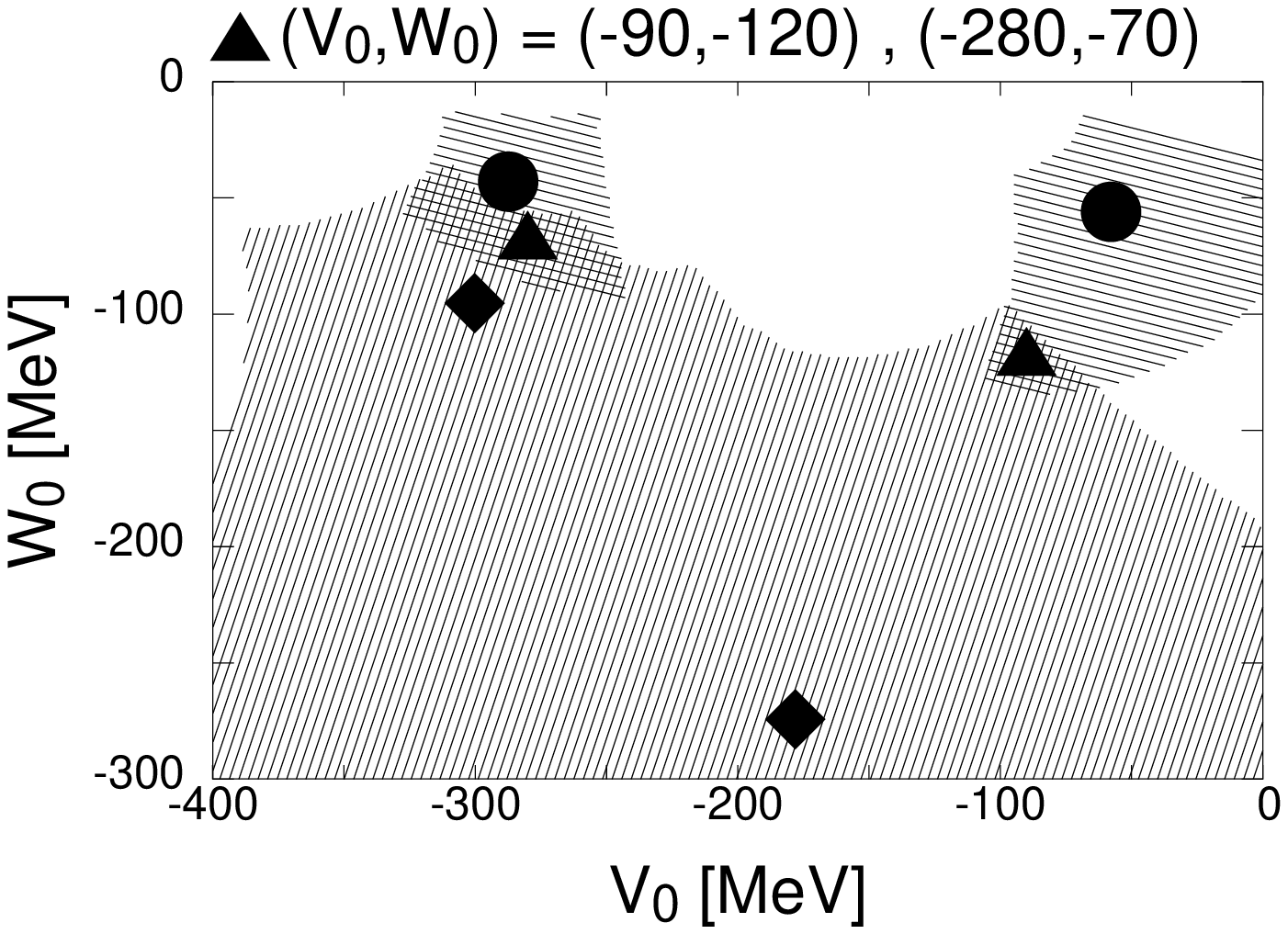}
\caption{The overlaid plot of the figures Fig.~\ref{Fig:1} (c) and Fig.~\ref{Fig:2} (c) is shown. The solid triangles show the two sets of potential parameters IS-A ($V_0,W_0)=(-90,-120)$ MeV and IS-B $(V_0,W_0)=(-280,-70)$ MeV that are consistent with both $K^--^3$He and $K^--^4$He atomic data in Ref.~\cite{J-PARCE62:2022qnt}.}
\label{Fig:3}
\end{figure}

We consider combined analyses of both data of $^3$He and $^4$He in this section.
First, we assume the isoscalar form (IS) for the optical potential, which does not distinguish the proton and neutron densities, and extract the potential parameters consistent with the data of $^3$He and $^4$He atoms simultaneously.
We show in Fig.~\ref{Fig:3} the overlaid contour plot of Fig.~\ref{Fig:1} (c) and Fig.~\ref{Fig:2} (c).
The potential parameters indicated by the solid triangles reproduce atomic states consistent with both $K^--^3$He and $K^--^4$He atom data.
These parameter sets can be categorized as weak-attraction strong-absorption potential ($V_0,W_0)=(-90,-120)$ MeV (IS-A) and strong-attraction weak-absorption potential ($V_0,W_0)=(-280,-70)$ MeV (IS-B).
We summarize the strengths of the potential IS-A and IS-B in Table~\ref{table:2}.
Note that because of the larger central densities of $^3$He and $^4$He than the normal nuclear density $\rho_0=0.17$ fm$^{-3}$, the depth of the potentials at the nuclear centers are larger than the potential strengths at $\rho_0$ expressed by the $(V_0,W_0)$ parameters.
These potential parameters are considered to reflect average behavior of the kaon interaction with He nuclei and can be compared with the potential parameters suited for global description of kaonic atoms. These potential parameters are applied further for heavier kaonic atoms in Appendix~\ref{sec:4}.

\begin{table}[!h]
\caption{The strengths of the isoscalar (IS) kaon-nucleus optical potential IS-A and IS-B, that reproduce all data~in Ref.~\cite{J-PARCE62:2022qnt} compiled in Table 1 within the errors for both $^3$He and $^4$He, are summarized for the phenomenological potential form defined in Eq.~(\ref{eq:1}).
The potential depths at normal nuclear density ($V_0$ and $W_0$ parameters) and those at the center of the $^3$He and $^4$He nuclei are shown for the potentials.
These potentials are corresponding to the solid triangles in Figs.~\ref{Fig:3},~\ref{Fig:11new},~\ref{Fig:12},~\ref{Fig:10}.}
\label{table:2}
\centering
\begin{tabular}{|c|c|c|c||c|c|}
\hline
\multicolumn{2}{|c|}{$^3$He and $^4$He}&\multicolumn{2}{c||}{IS-A}&\multicolumn{2}{c|}{IS-B}\\
%\cmidrule(rl){3-4}\cmidrule(lr){5-6}
\hline
\multicolumn{2}{|c|}{[MeV]}&Real&Imag&Real&Imag\\ \hline
\multicolumn{2}{|c|}{$U$ at $\rho=\rho_0$}&\multirow{2}{*}{$-90$}&\multirow{2}{*}{$-120$}&\multirow{2}{*}{$-280$}&\multirow{2}{*}{$-70$}\\
\multicolumn{2}{|c|}{($V_0$ and $W_0$)}&&&&\\
\hline
\multirow{4}{*}{$U(r=0)$}&\multirow{2}{*}{$^3$He}&\multirow{2}{*}{$-119$}&\multirow{2}{*}{$-159$}&\multirow{2}{*}{$-370$}&\multirow{2}{*}{$-93$}\\
&&&&&\\
\cline{2-6}
&\multirow{2}{*}{$^4$He}&\multirow{2}{*}{$-182$}&\multirow{2}{*}{$-243$}&\multirow{2}{*}{$-566$}&\multirow{2}{*}{$-142$}\\
&&&&&\\
\hline
\end{tabular}
\end{table}

We, then, consider the potential with the isospin dependence for the $K^--$He systems.
Actually, the potential strengths $V_0$ and $W_0$ for the kaonic $^3$He and $^4$He atoms listed in Table~\ref{table:1-5}
 and \ref{table:1-7} may suggest the different contributions from $K^-p$ and $K^-n$ interaction.
The isospin dependence of the ${\bar K}N$ interaction is also anticipated theoretically~\cite{Miyahara:2015bya}.
Though the accuracy of the data seems not enough to determine the isospin dependence, we consider, as an estimation, a phenomenological potential form with the isospin dependence as, 
\begin{equation}
\tilde{U}(r)=[V_p+iW_p]\frac{\rho_p(r)}{\rho_0}+[V_n+iW_n]\frac{\rho_n(r)}{\rho_0}~~,\label{eq:iso1}
\end{equation}
where $\rho_p$ and $\rho_n$ are the proton and neutron distributions.
Assuming the same form for the proton and neutron distributions for each nucleus, we have $\rho_p(r)=\displaystyle \frac{2}{3}\rho(r)$ and $\displaystyle \rho_n(r)=\frac{1}{3}\rho(r)$ for $^3$He and $\displaystyle \rho_p(r)=\rho_n(r)=\frac{1}{2}\rho(r)$ for $^4$He and, thus, the potential takes forms as,
\begin{equation}
\tilde{U}(r)=\frac{2V_p+V_n}{3}\frac{\rho(r)}{\rho_0}+i\frac{2W_p+W_n}{3}\frac{\rho(r)}{\rho_0}~~~~(\rm{for}~^3{\rm He})\label{eq:iso3}~,
\end{equation}
and 
\begin{equation}
\tilde{U}(r)=\frac{V_p+V_n}{2}\frac{\rho(r)}{\rho_0}+i\frac{W_p+W_n}{2}\frac{\rho(r)}{\rho_0}~~~~(\rm{for}~^4{\rm He})\label{eq:iso2}~.
\end{equation}
If we adopt these potential forms to the parameter sets $(V_0,W_0)=(-300,-95)$~MeV for $K^--^3$He and $(-287,-43)$~MeV for $K^--^4$He in Tables~\ref{table:1-5} and \ref{table:1-7}, we have~$(V_p,W_p)=(-326,-199)$~MeV, and ($V_n,W_n)=(-248,113)$~MeV.
The values for proton and neutron terms are largely different and they could indicate the existence of the relatively large isospin dependence in the kaon-nucleus optical potential.
At the same time the sign of $W_n$ is found to be positive, which is unnatural.
This might indicate that effects beyond linear density contributions, such as two nucleon absorptions of kaon, could be significant in such higher densities as reported in Ref.~\cite{Cieply:2011yz,Cieply:2011fy,Friedman:2016rfd,Hrtankova:2019jky,Obertova:2022wpw,Sekihara:2012wj}.
For further studies, we need more accurate data to determine the isospin dependence.

\subsection{Atomic $1s$ states in $K^--^3$He and $K^--^4$He}
In order to study further implications, we investigate the atomic $1s$ states in $^3$He and $^4$He.
The atomic $1s$ states are more sensitive to the strong interaction than the $2p$ state, because the $s$ states are free from the centrifugal barrier.
Thus, we expect to obtain further information on the optical potentials of He by experimental studies of the atomic $1s$ states and explore the validity of the potentials HE3-A, -B and HE4-A, -B.
Here we show our theoretical predictions of the binding energies and widths for the atomic $1s$ state in $^3$He and $^4$He calculated with the potentials determined by the experimental data of the $2p$ states.
With comparison of these calculations with future experiments we will be able to distinguish better potentials.

We adopt the potential parameters HE3-A, -B, and HE4-A, -B determined in Sect. 3.2 to the theoretical study of the $1s$ states.
The calculated results are shown in Table~\ref{table:4}.
The difference of the energy of the electromagnetic states in $^3$He and $^4$He mainly stems from the difference of the reduced mass of the systems.
Since the width of the $1s$ state is much smaller than the level spacing of the kaonic atom states, the $1s$ states will be able to find experimentally as a discrete state of deeply bound pionic atoms~\cite{Yamazaki:2012zza,Hirenzaki:2022dpt,Itahashi:2023boi,piAF:2022gvw}.
We also find clear differences in the predictions of the $1s$ states for two parameter sets both of $^3$He and $^4$He, although these parameters provide almost same energies and widths of the $2p$ states.
In the $K^--^3$He system, the calculation shows around $0.5$ keV difference in the shift of the $2p\to1s$ transition energy between the results by HE3-A and HE3-B.
The difference of the widths of the $1s$ state is so large that the width by HE3-B is more than twice of that by HE3-A.
In $K^--^4$He, the difference of the energy shift is larger than $3$ keV, while the difference in the $1s$ width is as small as around $0.6$ keV.
In this way, if one would observe the energy shift of the $1s$ state, one could distinguish the optical potentials providing an equivalent $2p$ state.

In the studies of the $1s$ states, the non-linear terms of the nuclear density $\rho(r)$ in the potential could change the binding energies and widths of the $1s$ states and may provide non-negligible effects to the kaon energies.
Here, we estimate the effects of the $\rho^2$ term in the potential to the binding energy and width of the $1s$ state using perturbation theory.
We calculate matrix elements of an optical potential with $\rho^2$ dependence in terms of the wavefunctions $\phi_{1s}$ of the $1s$ kaonic atom state in He as
\begin{eqnarray}
\Delta E&=&\int\phi^*_{1s}(V'+iW')\left(\frac{\rho(r)}{\rho_0}\right)^2\phi_{1s}d^3r\nonumber
\end{eqnarray}
where parameters $V'$ and $W'$  express strengths of the $\rho^2$ term in the optical potential at $\rho=\rho_0$.
We evaluate the overlap integral defined by 
\begin{equation}
I\equiv\int\phi_{1s}^*\left(\frac{\rho(r)}{\rho_0}\right)^2\phi_{1s}d^3r\nonumber
\end{equation}
and find $I=3.0\times 10^{-5}$~(HE3-A) and $2.2\times 10^{-4}$ (HE3-B) for $^3$He, and $I=3.5\times 10^{-4}$ (HE4-A) and $4.8\times 10^{-4}$ (HE4-B) for $^4$He, respectively, which are found to be about one order magnitude smaller than the corresponding overlap integral for the linear density.
The calculated results indicate that the $\rho^2(r)$ term in the optical potential with the strength of around several tens MeV~\cite{Sekihara:2012wj} at $\rho=\rho_0$ can shift the energies of $1s$ kaonic states in helium by several keV from the present results in Table~\ref{table:4}.
For the $1s$ state in $^3$He with HE3-A, the wavefunction $\phi_{1s}$ does not have a clear node inside nucleus and the value of integral is smaller than other cases.
For comparison, we perform similar estimation for $2p$ states, which provide $I=1.2\times10^{-8}$~(HE3-A) and $6.8\times10^{-8}$~(HE3-B) for $^3$He, and $I=9.2\times 10^{-8}$~(HE4-A) and $1.2\times 10^{-7}$~(HE4-B).
The obtained values of the overlap integral $I$ are 3-4 orders magnitude smaller than those of $1s$ states.
So, the effects of the $\rho^2$ term in the potential to $2p$ kaonic states would be order of 1 eV, or smaller.
This estimation is comparable to the theoretical value 0.3 eV reported in Ref.~\cite{Friedman:2011np}.

As for possible experiments for the $1s$ state observation, we could think of the X-ray spectroscopy as in Ref.~\cite{J-PARCE62:2022qnt} first as a realistic method.
As another possible candidate of the experiments, we could think of the $(K^-,p)$ reaction with nuclear targets that was studied theoretically in Ref.~\cite{Yamagata:2007cp}.
In this reaction one observes the energy of the emitted protons and obtains the energy spectrum of kaonic atoms by the missing mass technique.
We hope that our research will encourage to conduct experiments in the near future.

\begin{table}[!h]
\caption{The calculated energies ($E$) and widths ($\Gamma$) of the kaonic $1s$ and $2p$ atomic states for $^3$He and $^4$He in units of keV. The energies calculated only with the electromagnetic potential are shown in the column of EM. The results obtained with the full potential for parameters HE3-A, -B and HE4-A, -B are shown in the columns in the table as indicated.
The $2p\to 1s$ transition energies ($E_{2p\to 1s}$) and their shifts ($\Delta E$) due to the optical potential are also shown.}
\label{table:4}
\centering
\begin{tabular}{|c|c|c|c|c|c|}
\hline
$^3$He~~[keV]&EM&\multicolumn{2}{c|}{HE3-A}&\multicolumn{2}{c|}{HE3-B}\\
\hline
{\bf Atomic state}&$E$&$E$&$\Gamma$&$E$&$\Gamma$\\
\hline
$1s$&$-44.7484$&$-38.1513$&$5.6302$&$-38.6419$&$11.5738$\\
$2p$&$-11.1951$&$-11.1949$&$2.50\times 10^{-3}$&$-11.1949$&$2.49\times 10^{-3}$\\
\hline\hline
$E_{2p\to 1s}$&$33.5533$&\multicolumn{2}{c|}{$26.9564$}&\multicolumn{2}{c|}{$27.4470$}\\
\hline
$\Delta E$&&\multicolumn{2}{c|}{$-6.5969$}&\multicolumn{2}{c|}{$-6.1063$}\\
\hline
\end{tabular}
~\\~\\
\begin{tabular}{|c|c|c|c|c|c|}
\hline
$^4$He~~[keV]&EM&\multicolumn{2}{c|}{HE4-A}&\multicolumn{2}{c|}{HE4-B}\\
\hline
{\bf Atomic state}&$E$&$E$&$\Gamma$&$E$&$\Gamma$\\
\hline
$1s$&$-46.5017$&$-40.1397$&$5.2594$&$-37.0823$&$4.6759$\\
$2p$&$-11.6242$&$-11.6244$&$1.01\times 10^{-3}$&$-11.6244$&$9.99\times 10^{-4}$\\
\hline\hline
$E_{2p\to 1s}$&$34.8775$&\multicolumn{2}{c|}{$28.5153$}&\multicolumn{2}{c|}{$25.4579$}\\
\hline
$\Delta E$&&\multicolumn{2}{c|}{$-6.0564$}&\multicolumn{2}{c|}{$-9.4196$}\\
\hline
\end{tabular}
\end{table}

\subsection{Possible effects from kaonic nuclei in $^{3,4}$He and structure of kaonic atoms}
If the optical potential is attractively strong enough, kaonic nuclear states, which are compact bound states induced by strong interaction, can be formed.
If atomic states have the same quantum numbers with such nuclear states, they may receive certain influence from the nuclear states~\cite{Gal:1996pr}, such as level repulsion induced by state mixing.
Here we consider the effects of possible nuclear states on the kaonic atom structure in $^{3}$He and $^{4}$He.
In the studies of the kaonic nuclear structure, the hadronic degrees of freedom and the energy dependence of the interaction are expected to be more important.
Nevertheless, because we focus on the nature of the atomic states, our study on kaonic nuclear states here should be considered to be rather qualitative.
The mutual influence between the atomic and nuclear states of kaon with the same angular momentum are interesting to be mentioned.
Here we just use the same optical potential also for the nuclear states.

\begin{figure}[!h]
\centering
\includegraphics[width=12cm]{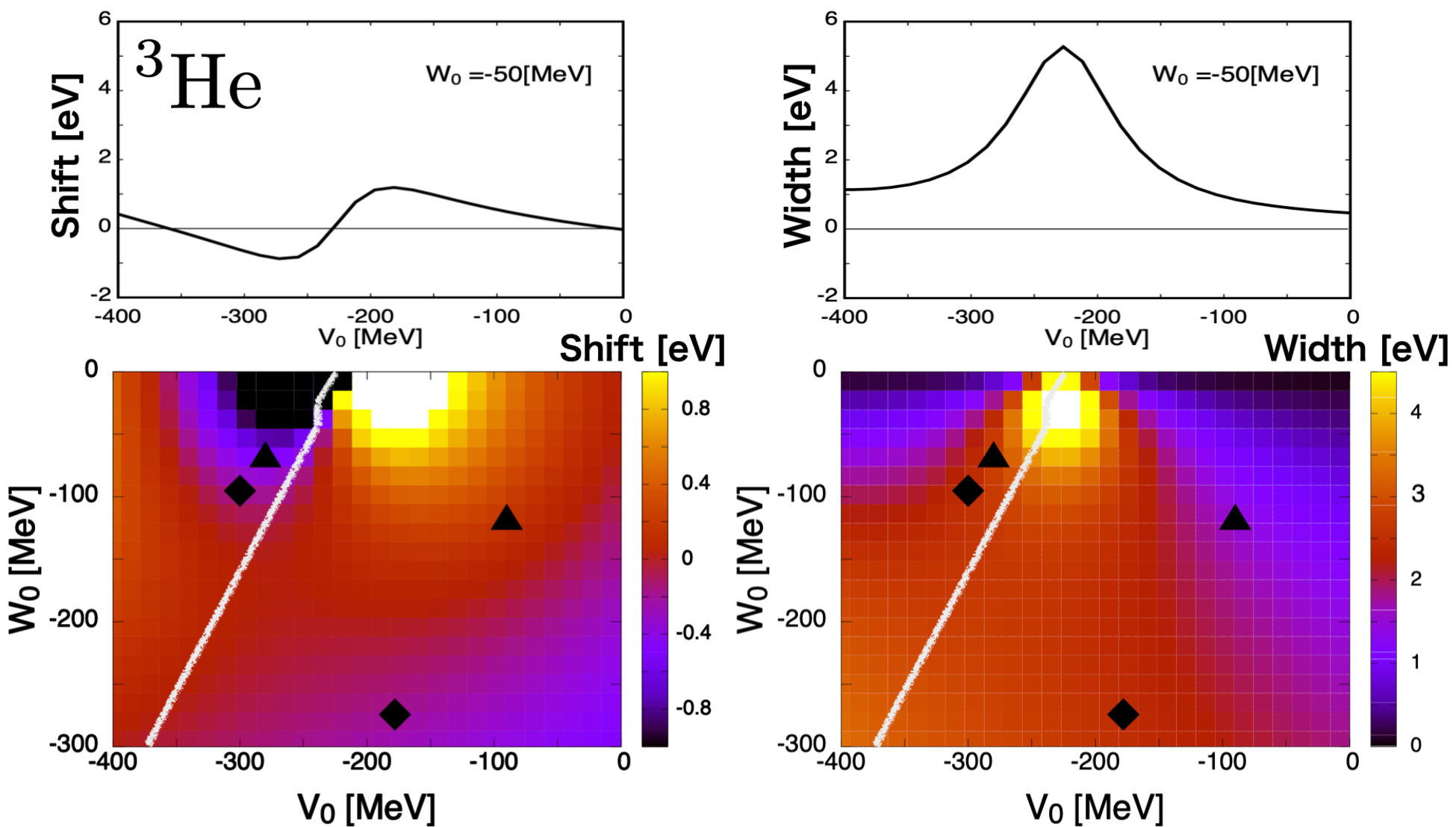}
     \caption{(Upper) The $V_0$ dependence of the shift of the $3d\to 2p$ transition energy and the $2p$ level width of the kaonic $^3$He atom calculated with $W_0=-50$ MeV. (Lower) the contour plots of the shift and the width of the kaonic $^3$He atom in the plane of the parameters $V_0$ and $W_0$.
The potential parameters in the region to the left of the white line allow the existence of the kaonic nuclear $p$ state(s). The solid diamonds in the figures indicate the potential parameter sets of HE3-A and HE3-B in Table~\ref{table:1-5} by which we can reproduce the experimental shift and width of the $K^--^3$He atom simultaneously.
     The solid triangles indicate the parameter sets of IS-A and IS-B in Table~\ref{table:2}.}
      \label{Fig:11new}
      \end{figure}

\begin{figure}[htbp]
\centering
\includegraphics[width=12cm,angle=-90]{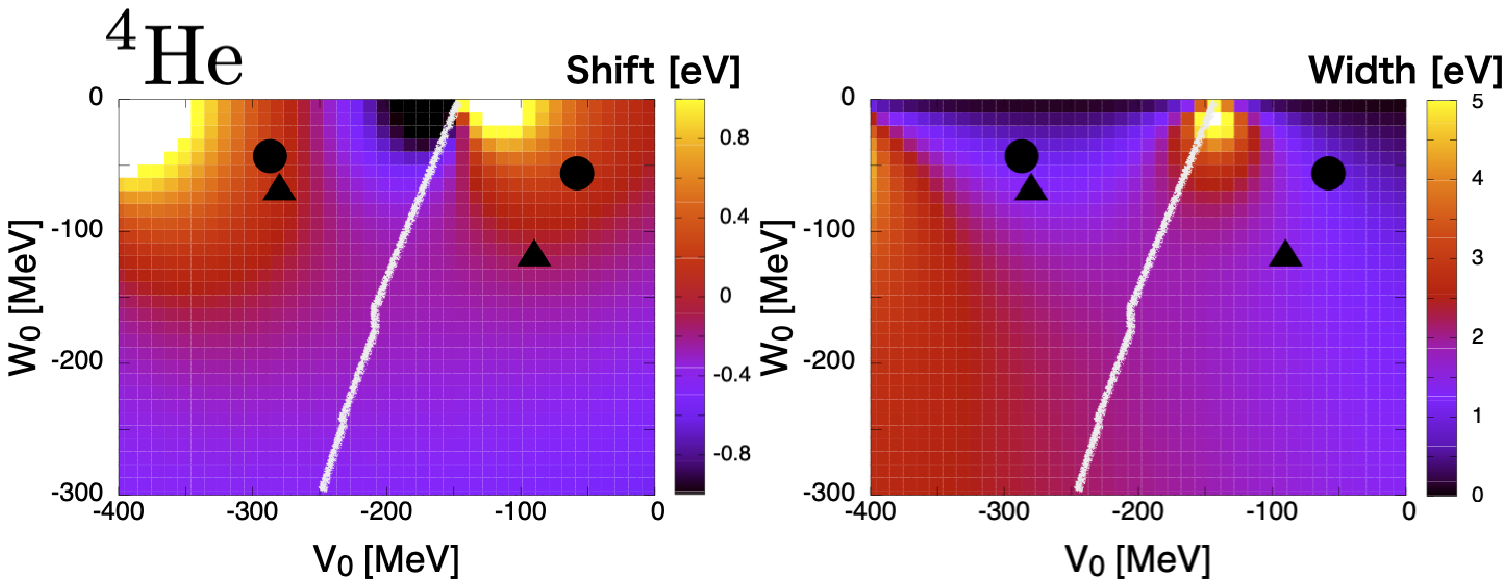}
      \caption{Same contour plots as Fig.~\ref{Fig:11new} (Lower) except for the kaonic $^4$He atomic state. The solid circles indicate the potential parameter sets of HE4-A and HE4-B in Table~\ref{table:1-7} by which we can reproduce the experimental shift and width of the $K^--^4$He atom simultaneously.}
      \label{Fig:12}
      \end{figure}

We show in Fig.~\ref{Fig:11new} (Lower) the contour plots of the shift and width of the $2p$ atomic state of the $K^--^3$He system in the $V_0-W_0$ plane, which are essentially the same plots as Fig.~\ref{Fig:1} but contain more detailed information.
In Fig.~\ref{Fig:11new}, we show the boundary of the potential parameters which allow the existence of the kaonic nuclear $p$ states by the white line and the nuclear $p$ state(s) exists in the parameter region left to the white line.
The shifts and widths obtained with smaller $W_0$ have strong $V_0$ dependence.
In particular, closed to the white line the shift suddenly changes its sign and the width has a maximum.
This feature is clearly shown in the upper figures of Fig.~\ref{Fig:11new} where the shift and width are plotted as functions of $V_0$ for a fixed $W_0=-50$ MeV.
These plots indicate typical influence of the existence of a nuclear state on the shift and width of the atomic state with the same angular momentum.
These features were studied in Ref.~\cite{Gal:1996pr} for bound states of negatively charged hadrons.
Closed to the white line which is the boundary for the existence of the nuclear state, the binding energy of the nuclear state is so small that large mixing of the atomic and nuclear states with the same quantum numbers take place.
These characteristic behavior of the shift and width stem from the occurrence of the level crossing between the nuclear state and atomic states. 
In Fig.~\ref{Fig:12}, we show the same contour plots as Fig.~\ref{Fig:11new} (Lower) for the atomic $2p$ state in $K^--^4$He.
The boundary of the existence of the nuclear $p$ state is again represented by the white line.
As we can see from Figs.~\ref{Fig:11new} and \ref{Fig:12}, two kinds of the potential parameter sets in Tables~\ref{table:1-5}, \ref{table:1-7}, and \ref{table:2} determined by the data~\cite{J-PARCE62:2022qnt} are distinguished by the boundaries of the existence of the kaonic nuclear $p$ states.

\begin{figure}[htbp]
\hspace{-1.0cm}
\centering
\includegraphics[width=12cm,angle=-90]{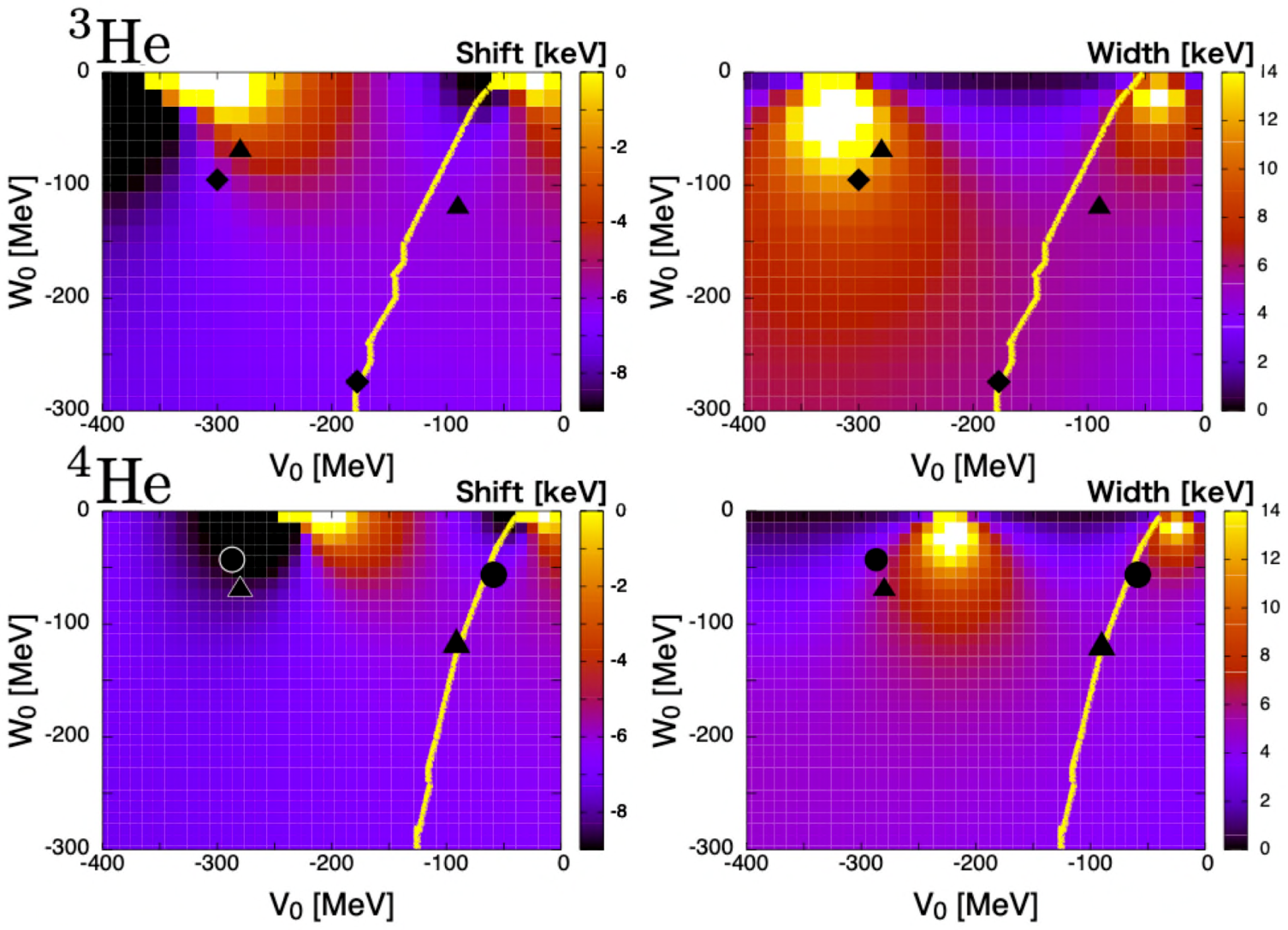}
\caption{The contour plots of the shifts of the $2p\to 1s$ transition energies (Left) and the $1s$ level widths (Right) of kaonic $^3$He (Upper) and $^4$He (Lower) atoms.
The solid triangles show the two sets of the potential parameters IS-A and IS-B in Table~\ref{table:2}.
The solid diamonds and circles indicate the potential parameters listed in Table~\ref{table:1-5} for $^3$He (HE3-A and HE3-B) and Table~\ref{table:1-7} for $^4$He (HE4-A and HE4-B), respectively.
The potential parameters in the region to the left of the yellow line allow the existence of the kaonic nuclear $s$ state(s).}
\label{Fig:10}
  \end{figure}

We show in Fig.~\ref{Fig:10} the contour plots of the shifts of the $2p\to 1s$ transition energies and the widths of the $1s$ states of kaonic atoms in $^3$He and $^4$He.
The parameter sets, IS-A and IS-B, are also indicated in the figures.
We can see the similar behavior of the shift and width for smaller $|W_0|$ region even with small $|V_0|$ to those in the contour plots of Fig.~\ref{Fig:11new} and \ref{Fig:12}.
This indicates the existence of the nuclear $s$ states because this characteristic behavior of the shift and width stems from the mixing of the atomic and nuclear states.
We show the region of the potential parameters where we can find the nuclear $s$ state(s) numerically
by the yellow line in Fig.~\ref{Fig:10} for $^3$He and $^4$He.
Considering also the results by E15 \cite{J-PARCE15:2018zys} and the tendency of the mass number dependence of the binding energy of the kaonic nuclear $s$ states found theoretically in Refs. \cite{Akaishi:2002bg,Yamazaki:2002uh,Shevchenko:2007zz,Wycech:2008wf,Dote:2008hw,Ikeda:2008ub,Barnea:2012qa,Sekihara:2016vyd,Ohnishi:2017uni,Dote:2017wkk}, we conclude that the existence of the nuclear $s$ state(s) in $^3$He and $^4$He is highly plausible.
As for the potential parameter sets reported in this article, the nuclear $s$ state(s) exists for parameter sets
IS-B for $^3$He and $^4$He, HE3-B for $^3$He, and HE4-B for $^4$He as clearly seen in Fig.~\ref{Fig:10}.
It is also clearly seen in the figure that the parameter set IS-A does not provide nuclear $s$ state in $^3$He.
For other cases, the parameters sets are just on the yellow line in Fig.~\ref{Fig:10} and the numerical calculations are a bit delicate, and it is difficult to make judgements regarding the existence of the nuclear $s$ state.
In our numerical calculations, we conclude the kaonic nuclear $s$ state exists for IS-A in $^4$He, and HE3-A in $^3$He.
On the other hand, the nuclear state cannot be found numerically for HE4-A in $^4$He. 
These results are rather qualitative, because for more precise calculations of nuclear states one would need further theoretical considerations for the optical potential such as energy dependence and few-body treatment.
The nuclear $s$ states can strongly affect the atomic $s$ states and change the binding energy and the width of the atomic $s$ states largely.
Thus, the origin of the discrepancies in the calculated results for the atomic $1s$ states by HE3-A, -B for $^3$He and HE4-A, -B for $^4$He in Table~\ref{table:4} is naturally expected to be related to the structure of the kaonic nuclear $s$ states.
Hence, the observation of the atomic $1s$ state in $^{3,4}$He is expected to provide the information on the potential parameters and the kaonic nuclear $s$ states, too.
At the same time, it will be interesting to fix the angular momentum of the kaonic nulcear state(s) in $^3$He and $^4$He, and clarify the (non-) existence of the kaonic nuclear $p$ state to develop the unified understandings of the structure of kaonic atoms and kaonic nuclei in He.

\section{Summary}
We study the features and the implications of the latest high precision kaonic atom data in Ref.~\cite{J-PARCE62:2022qnt} by considering the kaon-nucleus optical potential of $^3$He and $^4$He determined by the high precision data.
We adopt the phenomenological form of the optical potential and find the complex potential parameters by the experimental data in Ref.~\cite{J-PARCE62:2022qnt}.
The energy shifts and widths of the $K^--^3$He data and the $K^--^4$He data are reproduced by the potential parameters shown in Table~\ref{table:1-5} (HE3-A and HE3-B) and Table~\ref{table:1-7} (HE4-A and HE4-B), respectively.
The parameters in Table~\ref{table:2} (IS-A and IS-B) provide the energy shifts and widths consistent with the both data of $K^--^3$He and $K^--^4$He.
These potential parameters are consistent with the previous global analyses.
The optical potential obtained here may provide basic information for microscopic study of kaon$-$ He interactions.
We also find that the obtained parameters in Table~\ref{table:1-5} and Table~\ref{table:1-7} may indicate the possible strong isospin dependence of the kaon-nucleus optical potential, and we also find that the constraints to the potential parameters by the data of $^3$He are weaker than those by the data of $^4$He.
The comparison of the data with the results by the theoretical potentials~\cite{Ramos:1999ku,Friedman:2011np} are also shown.
%In Appendix, we show some additional calculated results using the potential parameters obtained in this article to see the implications of the potentials to other kaonic states.

Based on the present study of the kaonic He atoms, we can obtain the candidates of the optical potential parameters for individual nucleus thanks to the high precision experimental data.
The improvements of the data of the kaonic atom in $^3$He are found to be most effective for the   better determination of the interaction as seen in Section 3.
%Based on the present study of the kaonic He atoms, we think that the improvements of the data of kaonic $2p$ atoms in He is important for the better determination of the kaon-nucleus interaction as seen in Section 3.
We think that the further studies of the kaonic atoms especially $1s$ states in $^{3,4}$He are interesting and important to determine the kaon-nucleus interaction of helium isotopes decisively.
% and to develop the studies of the kaonic atoms as shown in Appendix.
We could even study the possible differences of the $K^--$ nucleus interaction between the $1s$ and $2p$ atomic states since all nucleons in a helium nucleus are in the $s$ state in a good approximation and thus, the angular momenta between $K^--$ nucleus and $K^--$ nucleon are expected to be same.
Hence, the atomic $1s$ state is expected to be more sensitive to ${\bar K}N$ $s$ wave amplitude than the atomic $2p$ state.

\section*{Acknowledgement}
This work was partly supported by JSPS KAKENHI Grant Numbers JP22K03607, JP23K03417.
The work of D.J. was partly supported by Grants-in-Aid for Scientific Research from JSPS (21K03530, 22H04917, 23K03427).

\appendix

\section{Implications of the optical potential from $K^--^{3,4}$He for heavier nuclei\label{sec:4}}
In this Appendix, we show our calculations of kaon bound states for heavier nuclei by using the optical potentials given in Table~\ref{table:2}. 

%In this Appendix, we apply the kaon-nucleus optical potential with the parameters in Tables~\ref{table:1-5}, \ref{table:1-7}, and \ref{table:2} to kaon bound states in heavier nuclei.
%, and investigate the features and implications of the optical potential determined from $K^--^{3,4}$He data.

We apply the isoscalar form of the optical potentials with the parameter sets IS-A and IS-B in Table~\ref{table:2} to heavier kaonic atoms.
We compare our calculated energy shifts and widths with the data in Refs.~\cite{Batty:1976ue,Backenstoss:1972yg,Wiegand:1973ac,Barnes:1974iu,Batty:1979zr,Kunselman:1971ib,Batty:1981dy,Miller:1975,Cheng:1975bj}, which are summarized in Table \ref{table:3}, and we find out the candidate of the potential parameters better suited for the kaonic atoms in other nuclei.
The potential suited for other nuclei is expected to describe well the average behavior of the kaon-nucleus interaction.
In the following calculations, we consider isotope nuclei having the largest natural abundance for each atomic number.
We use the charge distributions of the modified harmonic oscillator form for the nuclei with the atomic number $3\leq Z\leq 8$ and the Woods-Saxon form for those with larger $Z$.
The parameters of the charge distributions are taken from Refs.~\cite{DeJager:1974liz,DeVries:1987atn}.
The nuclear density distributions $\rho(r)$ in Eq.~(\ref{eq:1}) are deduced from the charge distributions as explained in Ref.~\cite{Nieves:1993ev}.

\begin{figure}[hbtp]
    \begin{tabular}{cc}
      \begin{minipage}[t]{0.5\hsize}
        \centering
        \includegraphics[keepaspectratio, scale=0.3,angle=-90]{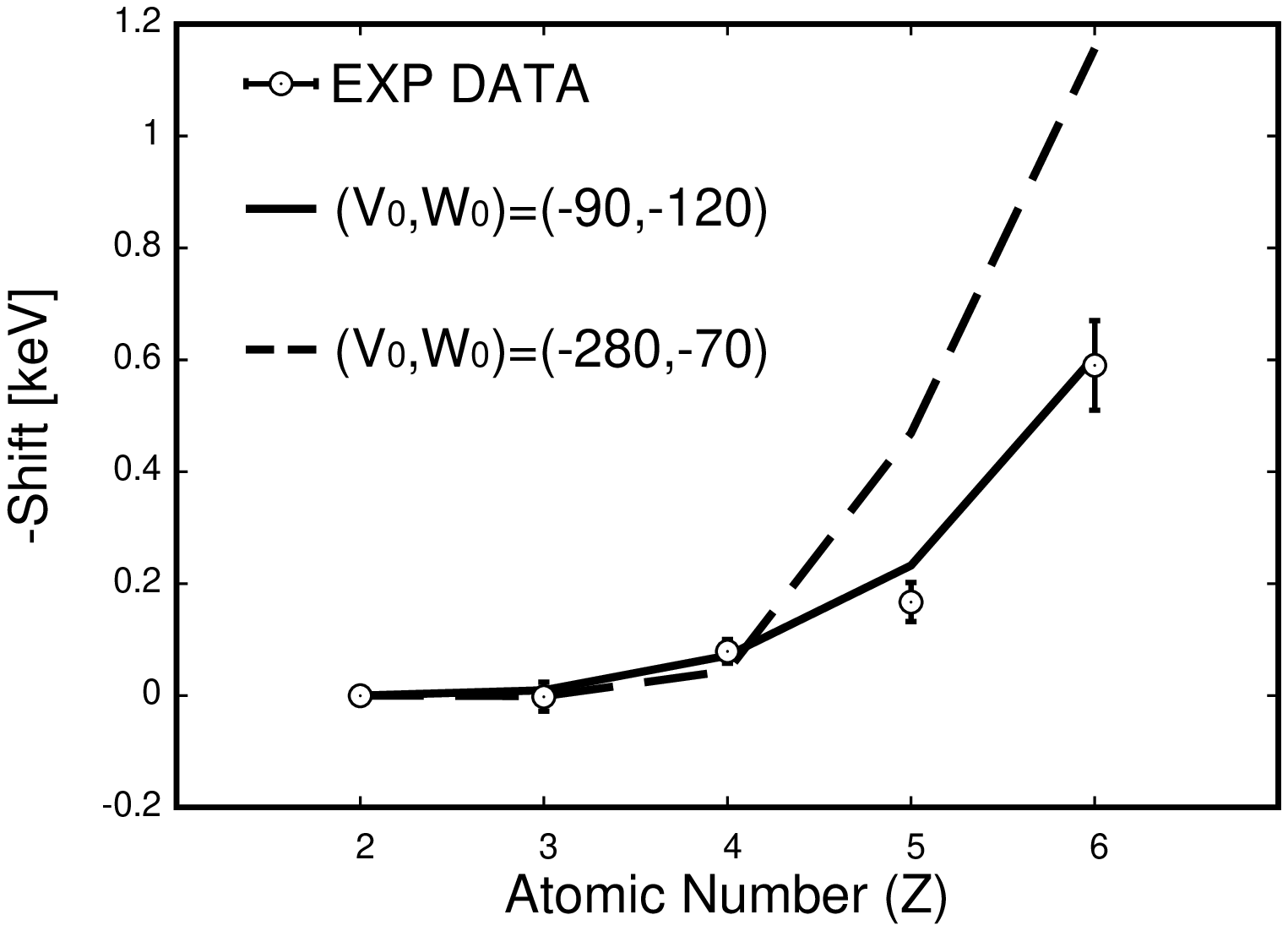}
%        \subcaption{Composite}
      \end{minipage} &
      \begin{minipage}[t]{0.5\hsize}
        \centering
        \includegraphics[keepaspectratio, scale=0.3,angle=-90]{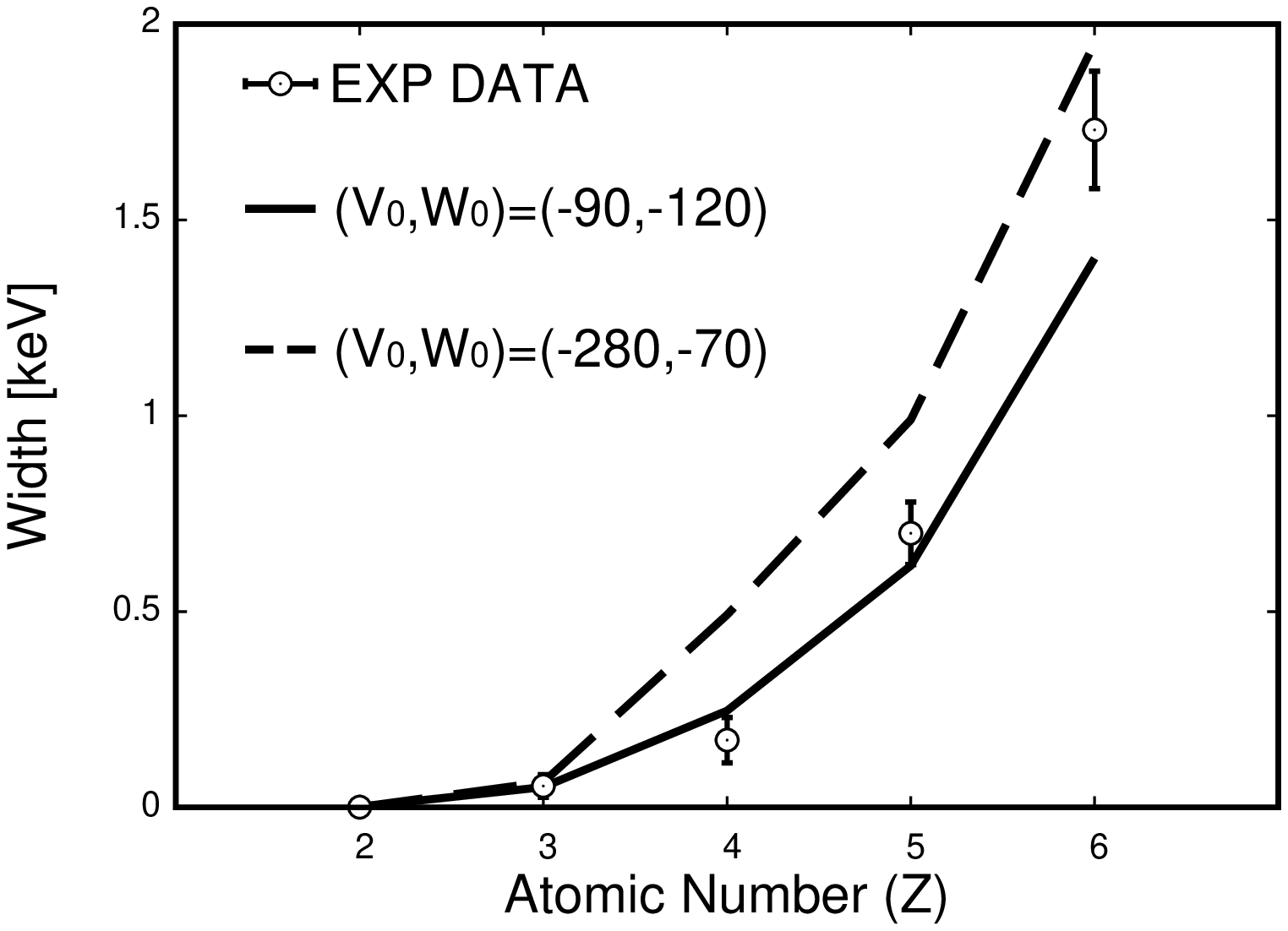}
  %      \subcaption{Gradation}
      \end{minipage} 
    \end{tabular}
     \caption{The calculated results of the energy shifts of the $3d\to 2p$ transition (Left) and the $2p$ level widths of the kaonic atoms (Right) with the experimental data compiled in Table~\ref{table:3}~\cite{Batty:1976ue,Backenstoss:1972yg,Wiegand:1973ac,Barnes:1974iu,Batty:1979zr,Kunselman:1971ib,Batty:1981dy,Miller:1975,Cheng:1975bj} . The horizontal axis shows the atomic number $Z$ of the nucleus. We consider an isotope nucleus with the largest natural abundance for each $Z$ for the theoretical calculations. For helium, we show the data of $^4$He in Ref.~\cite{J-PARCE62:2022qnt}. The parameters of the optical potential (IS-A and IS-B) used for the calculations are indicated in the figures. }
             \label{Fig:6}
  \end{figure}

\begin{figure}[htbp]
    \begin{tabular}{cc}
      \begin{minipage}[t]{0.5\hsize}
        \centering
        \includegraphics[keepaspectratio, scale=0.3,angle=-90]{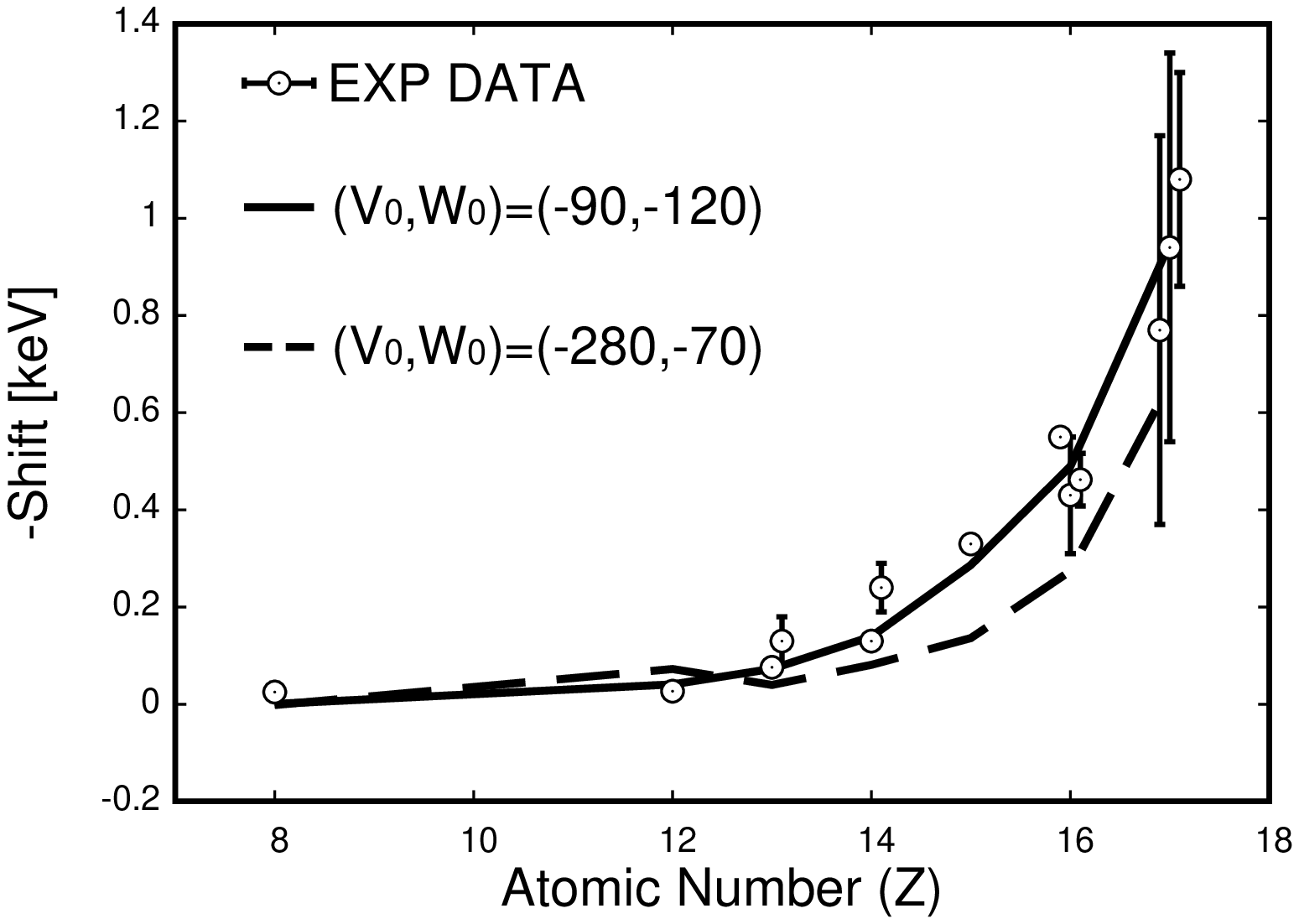}
%        \subcaption{Composite}
      \end{minipage} &
      \begin{minipage}[t]{0.5\hsize}
        \centering
        \includegraphics[keepaspectratio, scale=0.3,angle=-90]{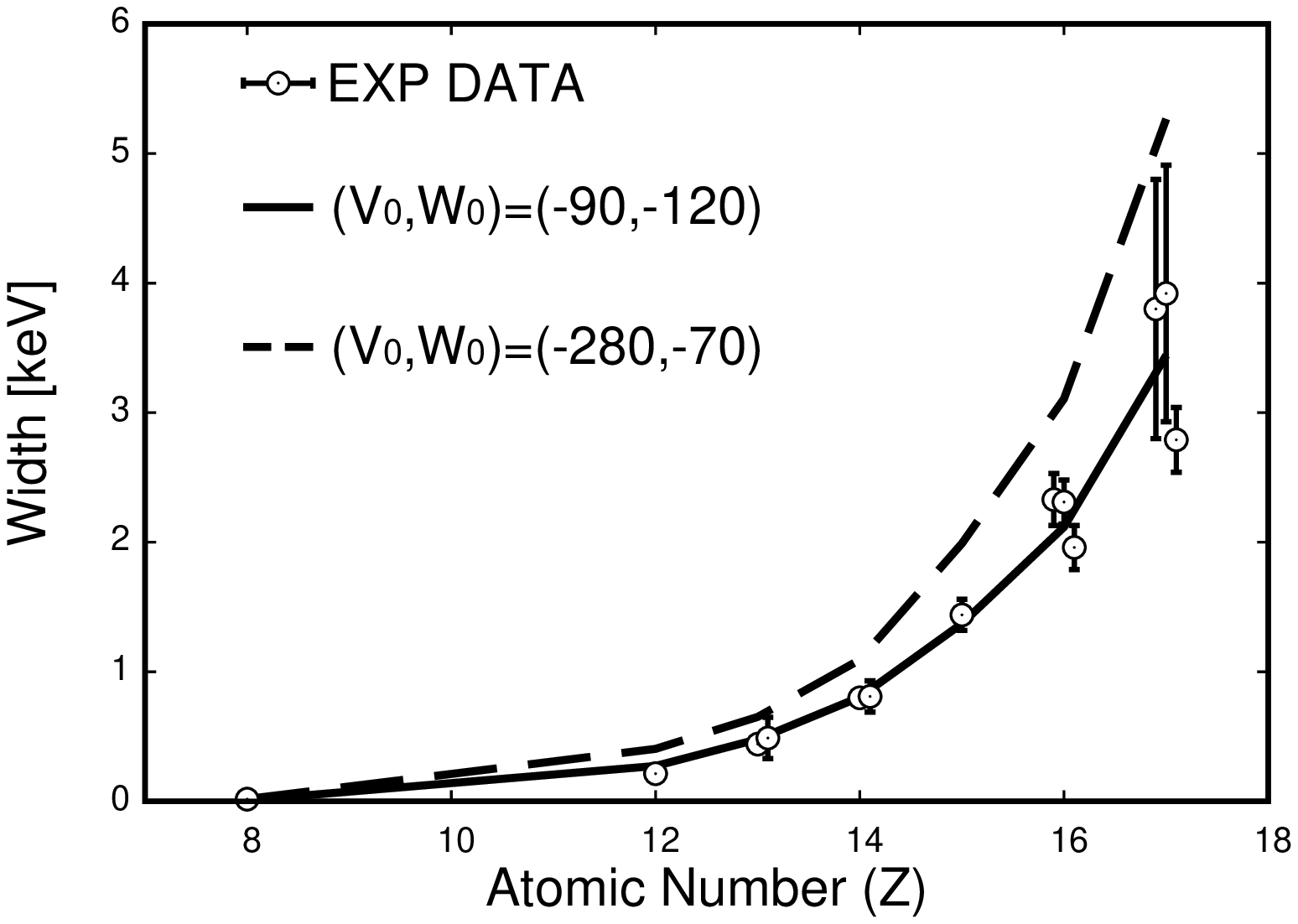}
  %      \subcaption{Gradation}
      \end{minipage} 
    \end{tabular}
     \caption{Same as Fig.~\ref{Fig:6} except for the $4f\to 3d$ transitions and the $3d$ level widths of the kaonic atoms.}
             \label{Fig:7}
  \end{figure}

\begin{figure}[htbp]
    \begin{tabular}{cc}
      \begin{minipage}[t]{0.5\hsize}
        \centering
        \includegraphics[keepaspectratio, scale=0.3,angle=-90]{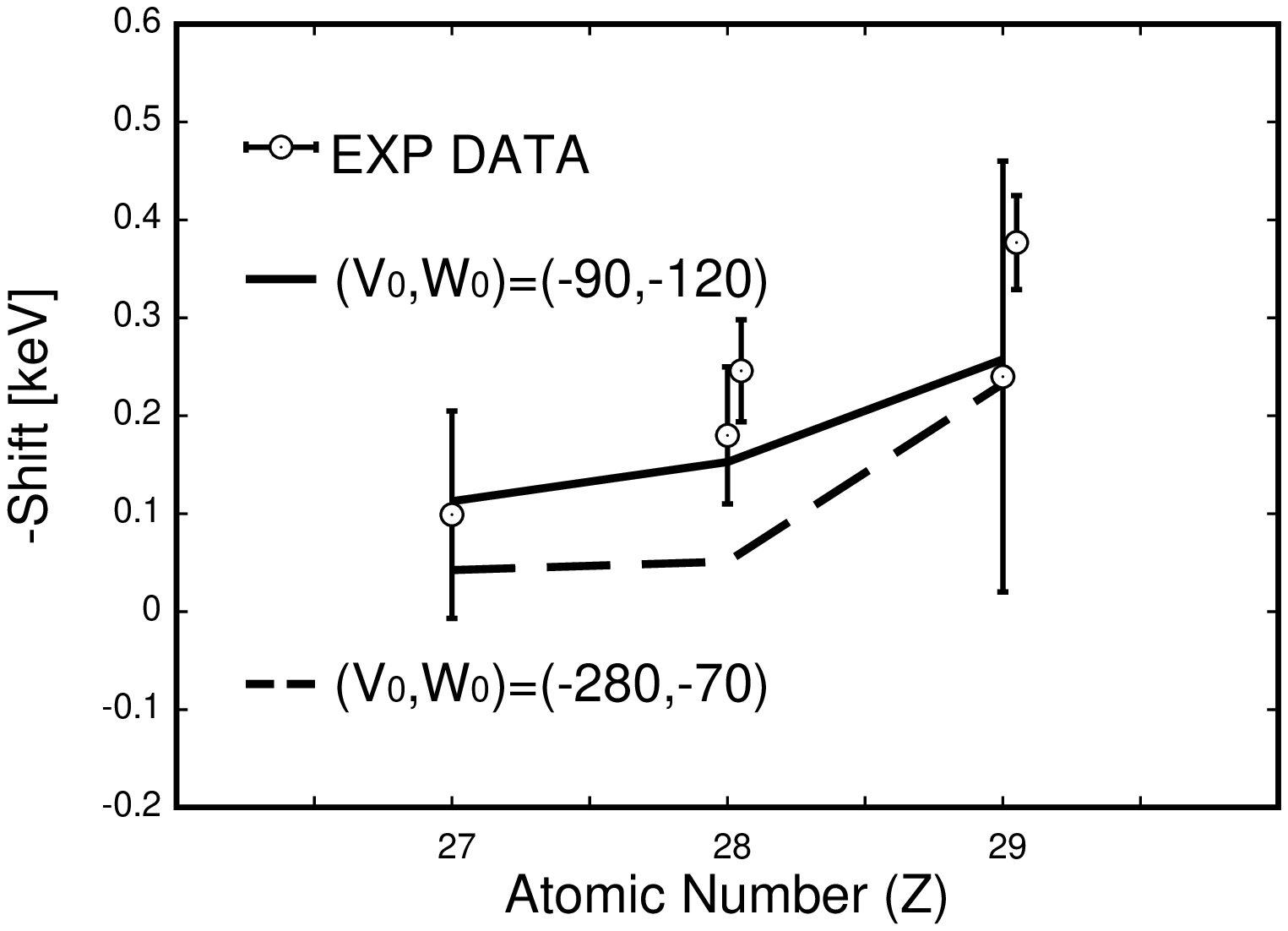}
%        \subcaption{Composite}
      \end{minipage} &
      \begin{minipage}[t]{0.5\hsize}
        \centering
        \includegraphics[keepaspectratio, scale=0.3,angle=-90]{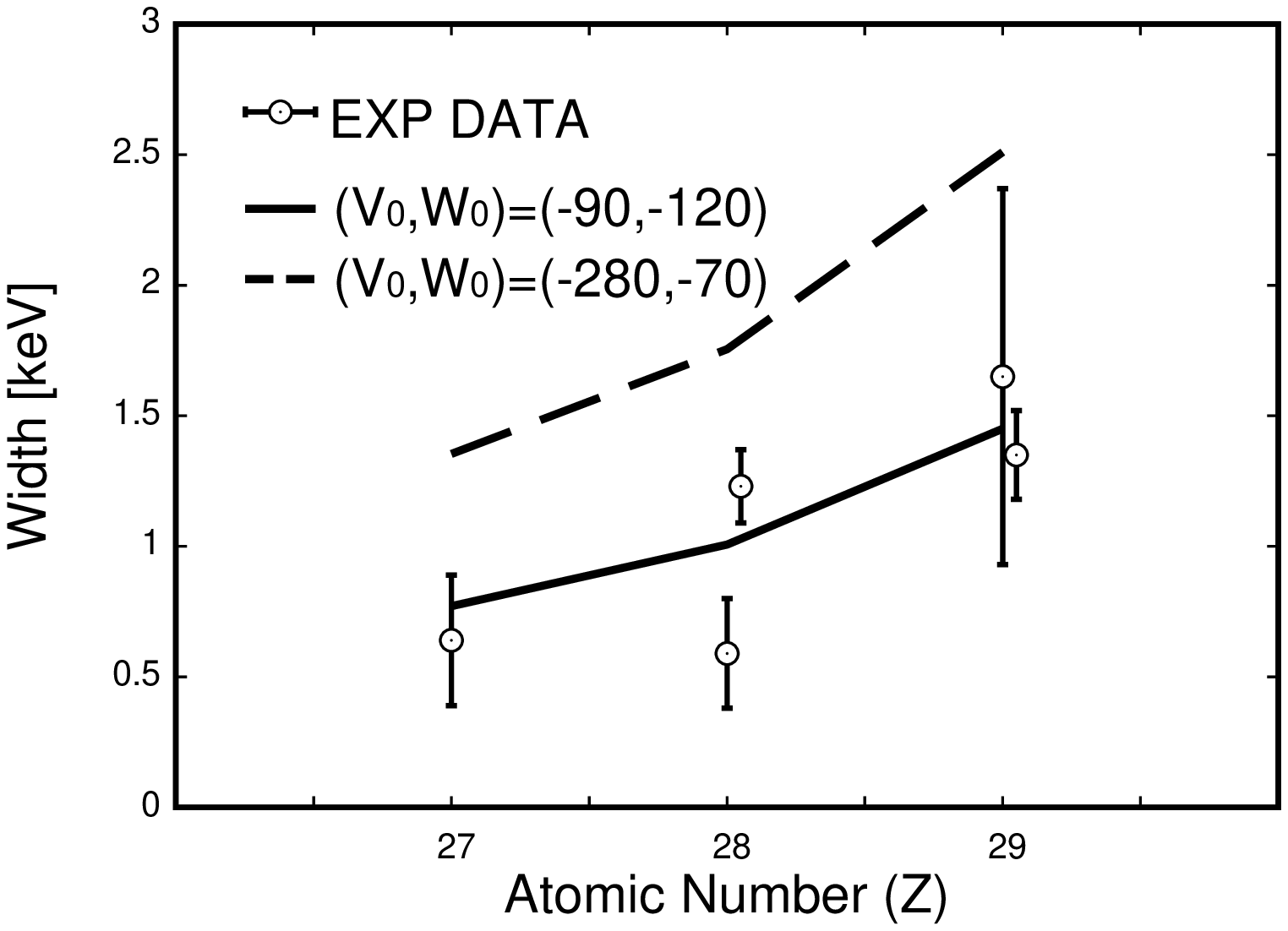}
  %      \subcaption{Gradation}
      \end{minipage} 
    \end{tabular}
     \caption{Same as Fig.~\ref{Fig:6} except for the $5g\to 4f$ transitions and the $4f$ level widths of the kaonic atoms.}
             \label{Fig:8}
  \end{figure}

\begin{figure}[htbp]
    \begin{tabular}{cc}
      \begin{minipage}[t]{0.5\hsize}
        \centering
        \includegraphics[keepaspectratio, scale=0.3,angle=-90]{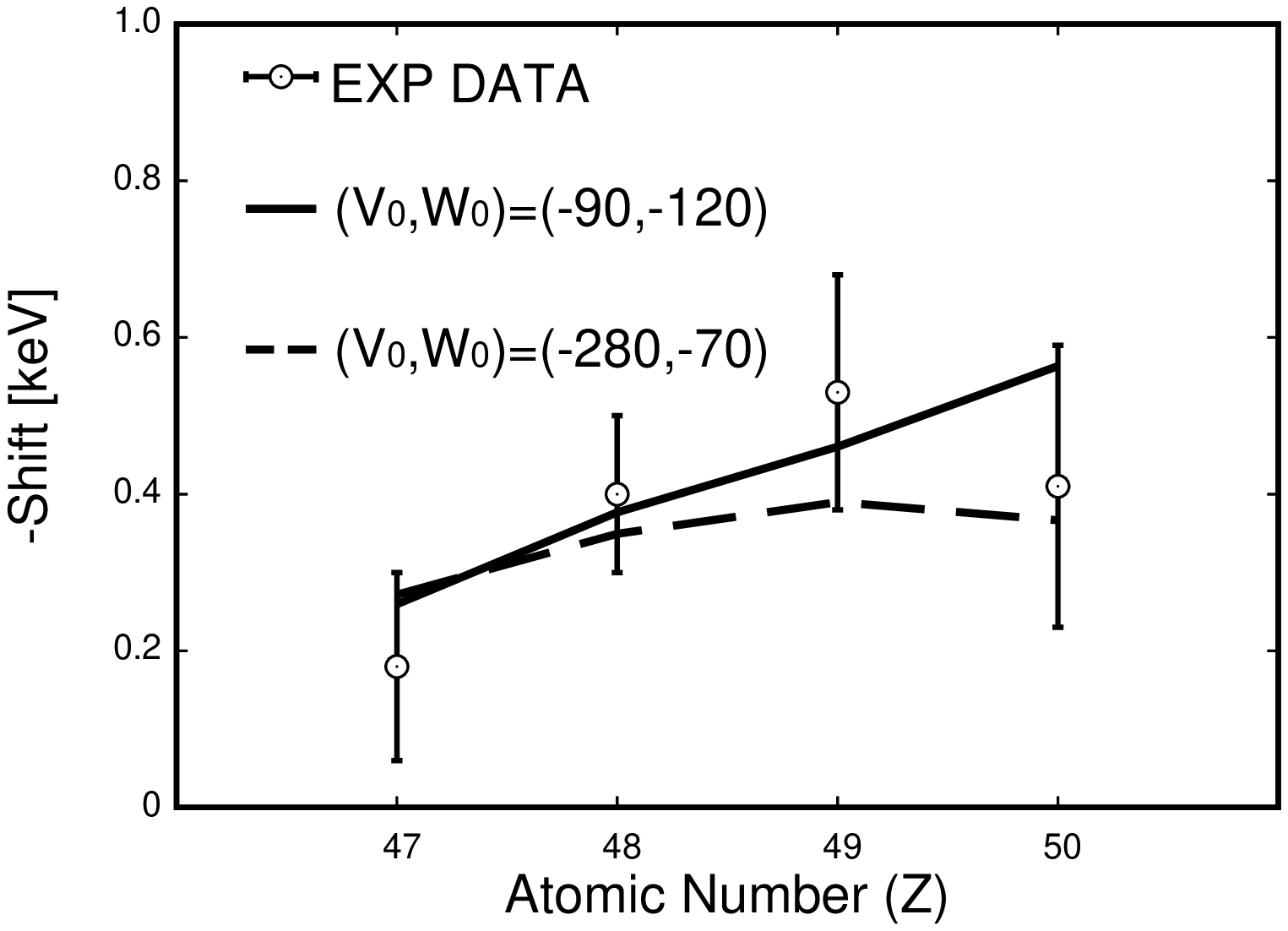}
%        \subcaption{Composite}
      \end{minipage} &
      \begin{minipage}[t]{0.5\hsize}
        \centering
        \includegraphics[keepaspectratio, scale=0.3,angle=-90]{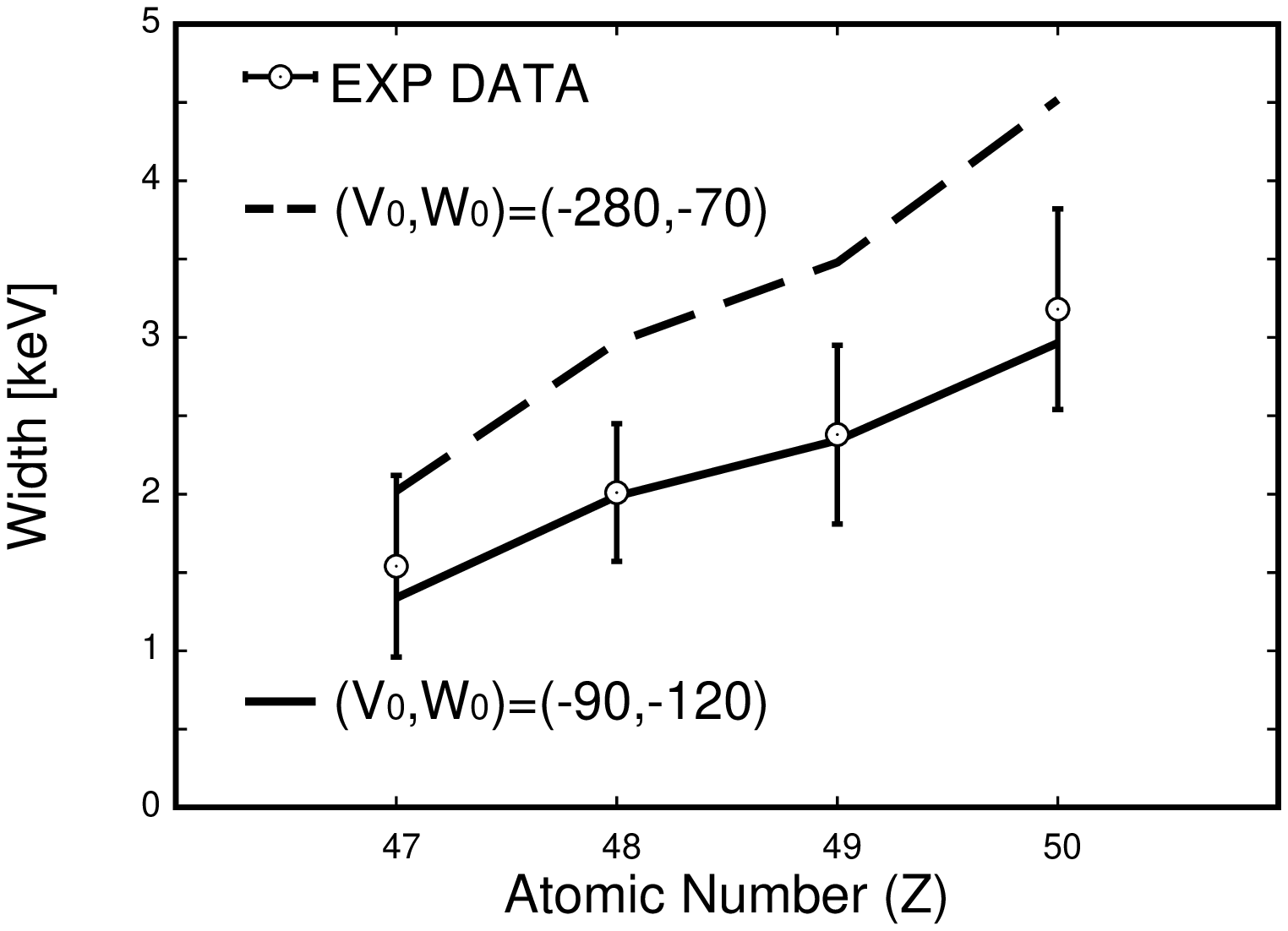}
  %      \subcaption{Gradation}
      \end{minipage} 
    \end{tabular}
     \caption{Same as Fig.~\ref{Fig:6} except for the $6h\to 5g$ transitions and the $5g$ level widths of the kaonic atoms.}
             \label{Fig:9}
  \end{figure}

\begin{figure}[htbp]
    \begin{tabular}{cc}
      \begin{minipage}[t]{0.5\hsize}
        \centering
        \includegraphics[keepaspectratio, scale=0.3,angle=-90]{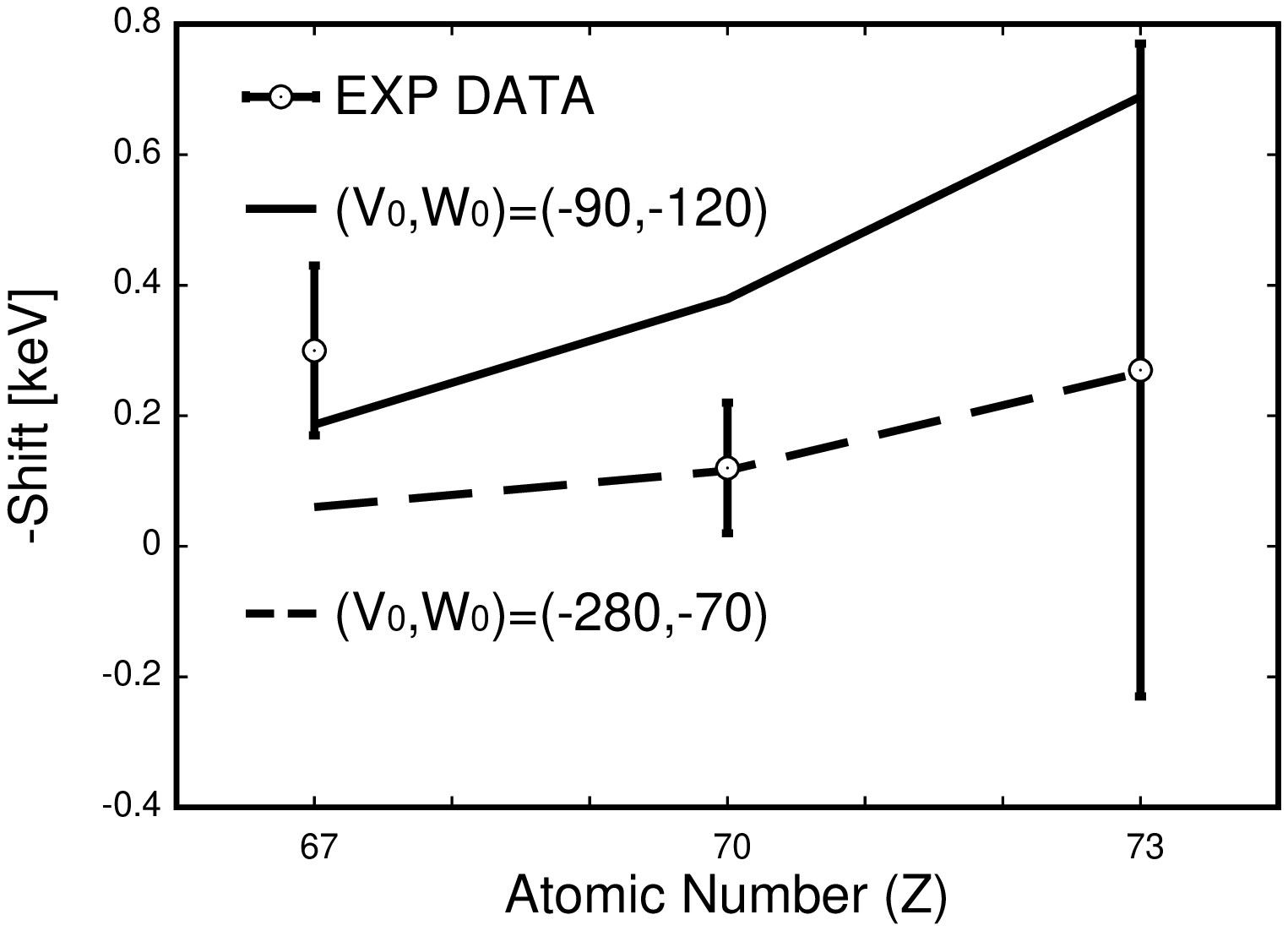}
%        \subcaption{Composite}
      \end{minipage} &
      \begin{minipage}[t]{0.5\hsize}
        \centering
        \includegraphics[keepaspectratio, scale=0.3,angle=-90]{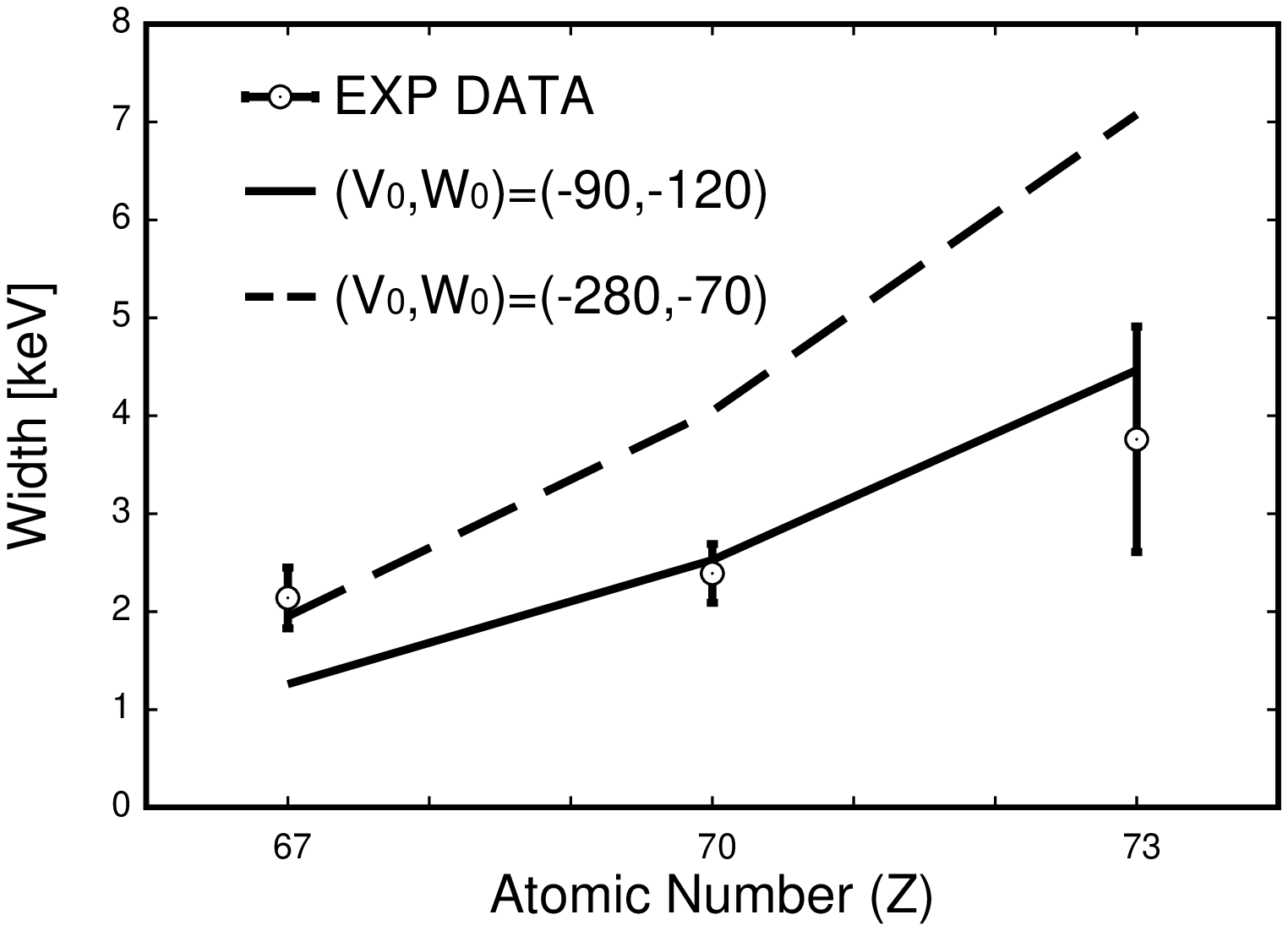}
  %      \subcaption{Gradation}
      \end{minipage} 
    \end{tabular}
     \caption{Same as Fig.~\ref{Fig:6} except for the $7i\to 6h$ transitions and the $6h$ level widths of the kaonic atoms.}
             \label{Fig:67}
  \end{figure}

\begin{figure}[htbp]
    \begin{tabular}{cc}
      \begin{minipage}[t]{0.5\hsize}
        \centering
        \includegraphics[keepaspectratio, scale=0.3,angle=-90]{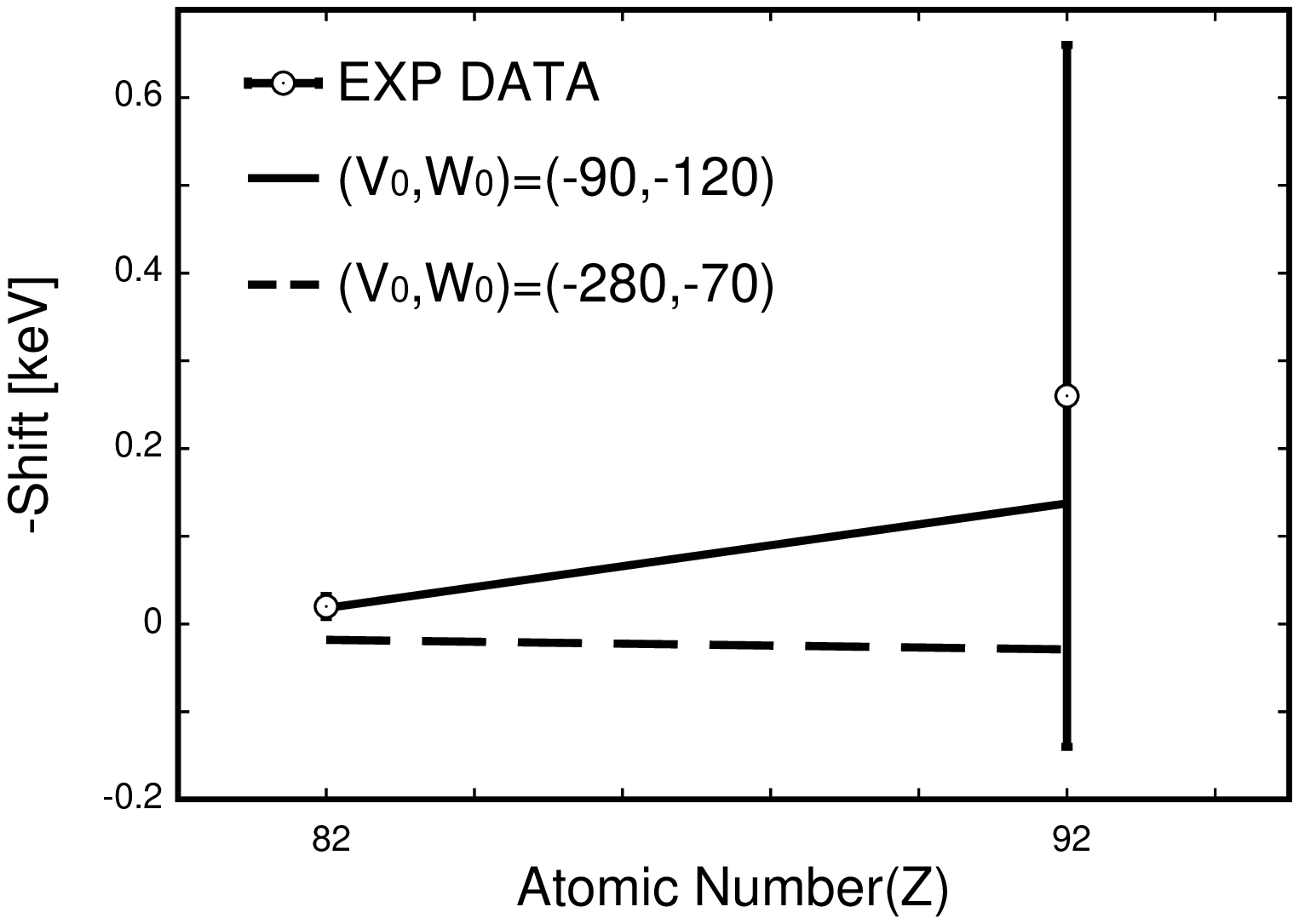}
%        \subcaption{Composite}
      \end{minipage} &
      \begin{minipage}[t]{0.5\hsize}
        \centering
        \includegraphics[keepaspectratio, scale=0.3,angle=-90]{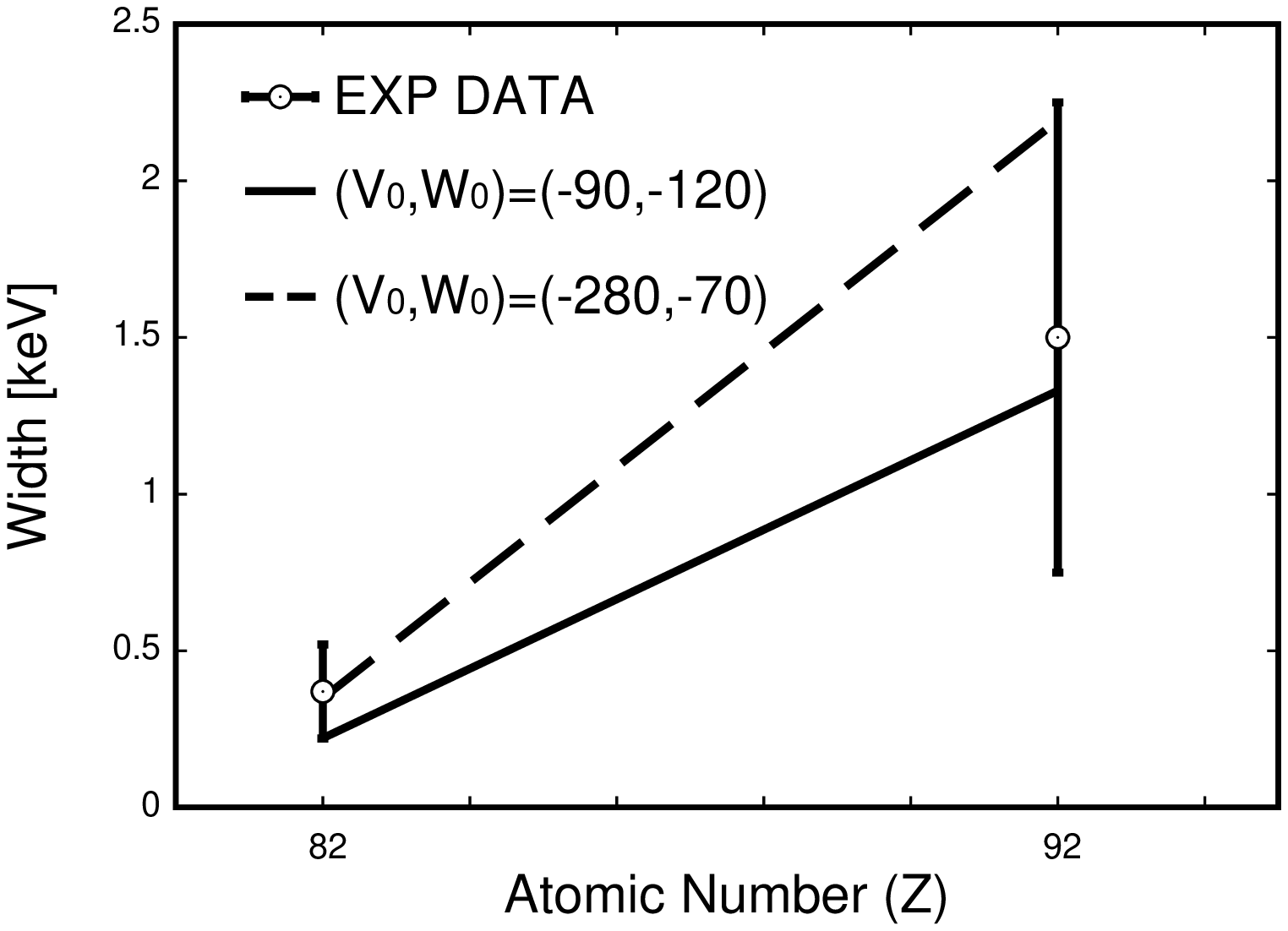}
  %      \subcaption{Gradation}
      \end{minipage} 
    \end{tabular}
     \caption{Same as Fig.~\ref{Fig:6} except for the $8k\to 7i$ transitions and the $7i$ level widths of the kaonic atoms.}
             \label{Fig:82}
  \end{figure}

The calculated results are shown and compared with the experimental data in Figs.~\ref{Fig:6}-\ref{Fig:82}.
\begin{table}[!h]
\caption{Observed data of the shift of the transition energies $\Delta E$ and the level widths $\Gamma$ for the heavier kaonic atoms considered in this appendix. The data are taken from Refs.~ \cite{Batty:1976ue,Backenstoss:1972yg,Wiegand:1973ac,Barnes:1974iu,Batty:1979zr,Kunselman:1971ib,Batty:1981dy,Miller:1975,Cheng:1975bj}.}
\label{table:3}
\centering
\begin{tabular}{c|c|c|c|c|c}
\hline
$Z$&Nucleus&Transition&$\Delta E$~[keV]&$\Gamma$~[keV]&Ref.\\
\hline
3&Li&$3d\to 2p$&$0.002\pm 0.026$&$0.055\pm 0.029$&\cite{Batty:1976ue}\\
4&Be&$3d\to 2p$&$-0.079\pm 0.021$&$0.172\pm 0.58$&\cite{Batty:1976ue}\\
5&B&$3d\to 2p$&$-0.167\pm0.035$&$0.700\pm 0.080$&\cite{Backenstoss:1972yg}\\
6&C&$3d\to 2p$&$-0.590\pm0.080$&$1.730\pm 0.150$&\cite{Backenstoss:1972yg}\\
8&O&$4f\to 3d$&$-0.025\pm 0.018$&$0.017\pm 0.014$&\cite{Batty:1979zr}\\
12&Mg&$4f\to 3d$&$-0.027\pm 0.015$&$0.214\pm 0.015$&\cite{Batty:1979zr}\\
13&Al&$4f\to 3d $&$-0.130\pm 0.050$&$0.490\pm 0.160$&\cite{Barnes:1974iu}\\
&&&$-0.076\pm 0.014$&$0.442\pm 0.022$&\cite{Batty:1979zr}\\
14&Si&$4f\to 3d$&$-0.240\pm 0.015$&$0.810\pm 0.120$&\cite{Barnes:1974iu}\\
&&&$-0.130\pm 0.015$&$0.800\pm 0.033$&\cite{Batty:1979zr}\\
15&P&$4f\to 3d$&$-0.330\pm 0.08$&$1.440\pm 0.120$&\cite{Backenstoss:1972yg}\\
16&S&$4f\to 3d $&$-0.550\pm0.06$&$2.330\pm 0.200$&\cite{Backenstoss:1972yg}\\
&&&$-0.43\pm0.12$&$2.310\pm0.170$&\cite{Wiegand:1973ac}\\
&&&$-0.462\pm0.054$&$1.96\pm0.17$&\cite{Batty:1979zr}\\
17&Cl&$4f\to 3d$&$-0.770\pm0.40$&$3.80\pm1.0$&\cite{Backenstoss:1972yg}\\
&&&$-0.94\pm0.40$&$3.92\pm0.99$&\cite{Kunselman:1971ib}\\
&&&$-1.08\pm0.22$&$2.79\pm0.25$&\cite{Wiegand:1973ac}\\
27&Co&$5g\to 4f$&$-0.099\pm 0.106$&$0.64\pm0.25$&\cite{Batty:1979zr}\\
28&Ni&$5g\to 4f$&$-0.180\pm0.070$&$0.59\pm0.21$&\cite{Barnes:1974iu}\\
&&&$-0.246\pm0.052$&$1.23\pm0.14$&\cite{Batty:1979zr}\\
29&Cu&$5g\to 4f$&$-0.240\pm0.220$&$1.650\pm0.72$&\cite{Barnes:1974iu}\\
&&&$-0.377\pm0.048$&$1.35\pm0.17$&\cite{Batty:1979zr}\\
47&Ag&$6h\to 5g$&$-0.18\pm0.12$&$1.54\pm0.58$&\cite{Batty:1979zr}\\
48&Cd&$6h\to 5g$&$-0.40\pm0.10$&$2.01\pm0.44$&\cite{Batty:1979zr}\\
49&In&$6h\to 5g$&$-0.53\pm0.15$&$2.38\pm0.57$&\cite{Batty:1979zr}\\
50&Sn&$6h\to 5g$&$-0.41\pm0.18$&$3.18\pm0.64$&\cite{Batty:1979zr}\\
67&Ho&$7i\to 6h$&$-0.30\pm0.13$&$2.14\pm0.31$&\cite{Batty:1981dy}\\
70&Yb&$7i\to 6h$&$-0.12\pm0.10$&$2.39\pm0.30$&\cite{Batty:1981dy}\\
73&Ta&$7i\to 6h$&$-0.27\pm0.50$&$3.76\pm1.15$&\cite{Batty:1981dy}\\
82&Pb&$8k\to7i$&$-$&$0.37\pm0.15$&\cite{Miller:1975}\\
&&&$-0.020\pm0.012$&$-$&\cite{Cheng:1975bj}\\
92&U&$8k\to7i$&$-0.26\pm0.4$&$1.50\pm0.75$&\cite{Miller:1975}\\
\hline
\end{tabular}
\end{table}
\noindent
As can be seen clearly in the figures, the parameter IS-A $(V_0,W_0)=(-90,-120)$~MeV is significantly better suited to describe the kaonic atom data in the wide range of the periodic table than IS-B.
The $\chi^2$ values calculated by the theoretical results of these potentials and the experimental data in Table~\ref{table:3} are found to be $\chi^2$(IS-A)$/$DoF$=91.1/62=1.5$ and $\chi^2$(IS-B)/DoF$=1014.7/62=16.4$, respectively.
The list of $\chi^2$ value for each datum is listed in Table~\ref{table:5}.
For comparison, we mention that the smallest value of $\chi^2$ obtained in Ref.~\cite{Obertova:2022wpw},
which include non-linear potential terms based on the $K^-N$ scattering amplitude derived from SU(3) chiral coupled channel models,
is $\chi^2$/DoF$=22.76/18 = 1.3$ for the eighteen observed data of kaonic atoms in nuclei from $^{12}$C to $^{208}$Pb. 
The potential with the parameter IS-B has a so large real part and provides the nuclear state with the same angular momenta as the atomic state as explained below.
Consequently, the level repulsion between atomic and nuclear states makes the shift of the atomic states repulsive, while the potential with the parameter IS-A has the large imaginary part which makes the shift repulsive.
Thus, the results obtained in this article, that the parameters IS-A is better suited for the global description of the kaonic atom data, agree with the results in Refs.~\cite{Batty:1997zp,Friedman:2007zza,Mares:2006vk,Iizawa:2019uiu} and show the same features of the kaon-nucleus optical potential which can reproduce the kaonic atom data in the wide range of the periodic table.
  
We also examine the possibility of the existence of kaonic nuclear states by using our isoscalar optical potential parameters IS-A and IS-B given in Table \ref{table:2} for heavier nuclei listed in Table \ref{table:3}, for which the energy  and width of the kaonic atomic states are observed.
We study the kaonic nuclear states that have the same angular momentum $\ell$ with the atomic state for each nucleus shown in Table \ref{table:3}, because these nuclear states can give influence to the atomic states observed in experiments due to the state mixing as discussed in Ref.~\cite{Gal:1996pr}.
We find that the optical potential IS-B, which has a larger attractive potential, provides kaonic nuclear states with the same angular momentum as the atomic state for all nucleus in Table \ref{table:3}, while the potential IS-A, which has a shallower attractive potential, does not give the corresponding nuclear states for nuclei with $Z\le28$ and provides bound states in Cu with $\ell = 3$ and heavier nuclei.
For the shallower IS-A potential, kaonic nuclear $p$ states exist for nuclei with $Z\geq29$.
These are much heavier than C of the heaviest nucleus in which the kaonic $2p$ atomic state is observed.

\begin{table}[!h]
\caption{The calculated $\chi^2$ values are shown for the experimental data listed in Table~\ref{table:3} and the theoretical results by the optical potential with the parameter sets IS-A and IS-B.
The sum of the $\chi^2$ values for the potential with IS-A and IS-B parameter sets are also shown in the last row.}
\label{table:5}
\centering
\begin{tabular}{c|c|c|c|c|c|c}
\hline
$Z$&Nucleus&Transition&\multicolumn{2}{c|}{IS-A}&\multicolumn{2}{c}{IS-B}\\%&Exp. Ref.\\
&&&$\Delta E$&$\Gamma$&$\Delta E$&$\Gamma$\\
\hline
3&Li&$3d\to 2p$&$0.199$&$0.009$&0.002&0.115\\%&\cite{Batty:1976ue}\\
4&Be&$3d\to 2p$&0.132&0.017&2.784&0.305\\%&\cite{Batty:1976ue}\\
5&B&$3d\to 2p$&0.231&3.248&20.206&24.275\\%&\cite{Backenstoss:1972yg}\\
6&C&$3d\to 2p$&3.517&1.063&74.509&13.143\\%&\cite{Backenstoss:1972yg}\\
8&O&$4f\to 3d$&0.042&4.779&50.105&2.185\\%&\cite{Batty:1979zr}\\
12&Mg&$4f\to 3d$&1.830&0.327&2.180&0.021\\%&\cite{Batty:1979zr}\\
13&Al&$4f\to 3d $&1.309&0.004&3.264&1.027\\%&\cite{Barnes:1974iu}\\
&&&0.856&16.077&9.216&163.901\\%&\cite{Batty:1979zr}\\
14&Si&$4f\to 3d$&4.056&0.002&10.078&5.560\\%&\cite{Barnes:1974iu}\\
&&&0.385&0.016&10.555&78.809\\%&\cite{Batty:1979zr}\\
15&P&$4f\to 3d$&0.299&0.318&5.884&21.091\\%&\cite{Backenstoss:1972yg}\\
16&S&$4f\to 3d $&1.015&1.129&21.153&15.115\\%&\cite{Backenstoss:1972yg}\\
&&&0.246&1.283&1.689&22.011\\%&\cite{Wiegand:1973ac}\\
&&&0.261&0.858&12.115&45.567\\%&\cite{Batty:1979zr}\\
17&Cl&$4f\to 3d$&0.201&0127&0.074&2.149\\%&\cite{Backenstoss:1972yg}\\
&&&0.001&0.232&0.487&1.849\\%&\cite{Kunselman:1971ib}\\
&&&0.353&6.827&3.630&98.093\\%&\cite{Wiegand:1973ac}\\
27&Co&$5g\to 4f$&0.017&0.275&0.285&8.161\\%&\cite{Batty:1979zr}\\
28&Ni&$5g\to 4f$&0.152&3.939&3.396&30.785\\%&\cite{Barnes:1974iu}\\
&&&3.216&2.542&14.062&14.072\\%&\cite{Batty:1979zr}\\
29&Cu&$5g\to 4f$&0.006&0.077&0.001&1.426\\%&\cite{Barnes:1974iu}\\
&&&6.184&0.349&9.026&46.537\\%&\cite{Batty:1979zr}\\
47&Ag&$6h\to 5g$&0.438&0.122&0.583&0.684\\%&\cite{Batty:1979zr}\\
48&Cd&$6h\to 5g$&0.125&0.422&0.097&1.433\\%&\cite{Batty:1979zr}\\
49&In&$6h\to 5g$&0.215&0.004&0.874&3.717\\%&\cite{Batty:1979zr}\\
50&Sn&$6h\to 5g$&0.726&0.113&0.058&4.371\\%&\cite{Batty:1979zr}\\
67&Ho&$7i\to 6h$&0.763&8.048&3.401&0.368\\%&\cite{Batty:1981dy}\\
70&Yb&$7i\to 6h$&6.707&0.208&0.002&30.539\\%&\cite{Batty:1981dy}\\
73&Ta&$7i\to 6h$&0.702&0.378&0.0001&8.320\\%&\cite{Batty:1981dy}\\
82&Pb&$8k\to7i$&$-$&0.984&$-$&0.020\\%&\cite{Miller:1975}\\
&&&0.022&$-$&9.988&$-$\\%&\cite{Cheng:1975bj}\\
92&U&$8k\to7i$&0.094&0.051&0.521&0.853\\%&\cite{Miller:1975}\\
\hline\hline
\multicolumn{3}{c}{$\chi^2$~total}&\multicolumn{2}{c}{91.082}&\multicolumn{2}{c}{1014.693}\\
\hline
\end{tabular}
\end{table}

\end{document}